\documentclass[11pt]{article}

\newsavebox{\foobox}
\newcommand{\slantbox}[2][0]{\mbox{%
        \sbox{\foobox}{#2}%
        \hskip\wd\foobox
        \pdfsave
        \pdfsetmatrix{1 0 #1 1}%
        \llap{\usebox{\foobox}}%
        \pdfrestore
}}
\newcommand\unslant[2][-.25]{\slantbox[#1]{$#2$}}

\newcommand{\mmu}{\text{\unslant\mu}}
\newcommand{\mpi}{\text{\unslant[-.18]\pi}}
\newcommand{\mdelta}{\text{\unslant[-.18]\delta}}

\usepackage[left=2cm, right=2cm, top=2.5cm, bottom=2.5cm]{geometry}
\usepackage[x11names]{xcolor}
\usepackage{fancyhdr, amssymb, cancel, amsmath, graphicx, pgfplots, tikz}
\usepackage{isomath}

\usetikzlibrary{shadows}

\newcommand{\stylecolor}{blue!50!black}

\usepackage[labelfont={bf,sf, color=\stylecolor}, margin={1.5cm,0cm}]{caption}

\usepackage[colorlinks=true, urlcolor=\stylecolor!70!white, linkcolor=\stylecolor, citecolor=\stylecolor!70!white, hyperindex=true, linktocpage=true]{hyperref}

\usepackage[explicit]{titlesec}

\newcommand*\sectionlabel{}
\titleformat{\section}
  {\gdef\sectionlabel{}
   \Large\bfseries\scshape}
  {\gdef\sectionlabel{\thesection }}{0pt}
  {\begin{tikzpicture}[remember picture,overlay]
       \end{tikzpicture}
  }
\titlespacing*{\section}{0pt}{0pt}{0pt}

\newcommand*\subsectionlabel{}
\titleformat{\subsection}
  {\gdef\subsectionlabel{}
   \large\bfseries\scshape}
  {\gdef\subsectionlabel{\thesubsection  }}{0pt}
  {\begin{tikzpicture}[remember picture,overlay]
    	\draw (-0.15, 0.02) node[right] {\color{\stylecolor} \textsf{\subsectionlabel}};
	\draw (1.25, 0) node[right] {\color{\stylecolor} \textsf{#1}};
	\fill[color=\stylecolor] (1,-0.25) rectangle (1.1, 0.25);
       \end{tikzpicture}
  }
\titlespacing*{\subsection}{0pt}{10pt}{10pt}

\newcommand*\subsubsectionlabel{}
\titleformat{\subsubsection}
  {\gdef\subsubsectionlabel{}
   \bfseries\scshape}
  {\gdef\subsubsectionlabel{\thesubsubsection.\ \  }}{0pt}
  {\begin{tikzpicture}[remember picture,overlay]
    	\draw (-0.15, 0) node[right] {\color{\stylecolor} \textsf{\subsubsectionlabel#1}};
       \end{tikzpicture}
  }
\titlespacing*{\subsubsection}{0pt}{7pt}{7pt}

\pgfplotsset{every axis legend/.append style={at={(1.02,1)},anchor=north west}}

\begin{document}

\allowdisplaybreaks

\pagestyle{fancy}
\renewcommand{\headrulewidth}{0pt}
\fancyhead{}

\fancyfoot{}
\fancyfoot[C] {\textsf{\textbf{\thepage}}}

\begin{equation*}
\begin{tikzpicture}
\draw (\textwidth, 0) node[text width = \textwidth, right] {\color{white} easter egg};
\end{tikzpicture}
\end{equation*}

\begin{equation*}
\begin{tikzpicture}
\draw (0.5\textwidth, -3) node[text width = \textwidth] {\huge  \textsf{\textbf{Sound waves and resonances in electron-hole plasma}} };
\end{tikzpicture}
\end{equation*}
\begin{equation*}
\begin{tikzpicture}
\draw (0.5\textwidth, 0.1) node[text width=\textwidth] {\large \color{black} \textsf{Andrew Lucas}};
\draw (0.5\textwidth, -0.5) node[text width=\textwidth] {\small\textsf{Department of Physics, Harvard University, Cambridge, MA 02138, USA}};
\end{tikzpicture}
\end{equation*}
\begin{equation*}
\begin{tikzpicture}
\draw (0, -13.1) node[right, text width=0.5\paperwidth] {\texttt{lucas@fas.harvard.edu}};
\draw (\textwidth, -13.1) node[left] {\textsf{\today}};
\end{tikzpicture}
\end{equation*}
\begin{equation*}
\begin{tikzpicture}
\draw[very thick, color=\stylecolor] (0.0\textwidth, -5.75) -- (0.99\textwidth, -5.75);
\draw (0.12\textwidth, -6.25) node[left] {\color{\stylecolor}  \textsf{\textbf{Abstract:}}};
\draw (0.53\textwidth, -6) node[below, text width=0.8\textwidth, text justified] {\small Inspired by the recent experimental signatures of relativistic hydrodynamics in graphene, we investigate theoretically the behavior of hydrodynamic sound modes in such quasi-relativistic fluids near charge neutrality, within linear response.   Locally driving an electron fluid at a resonant frequency to such a sound mode can lead to large increases in the electrical response at the edges of the sample, a signature which cannot be explained using diffusive models of transport.   We discuss the robustness of this signal to various effects, including electron-acoustic phonon coupling, disorder, and long-range Coulomb interactions.   These long range interactions convert the sound mode into a collective plasmonic mode at low frequencies unless the fluid is charge neutral.  At the smallest frequencies, the response in a disordered fluid is quantitatively what is predicted by a ``momentum relaxation time" approximation.   However, this approximation fails at higher frequencies (which can be parametrically small), where the classical localization of sound waves cannot be neglected.     Experimental observation of such resonances is a clear signature of relativistic hydrodynamics, and provides an upper bound on the viscosity of the electron-hole plasma.};
\end{tikzpicture}
\end{equation*}

\tableofcontents

\titleformat{\section}
  {\gdef\sectionlabel{}
   \Large\bfseries\scshape}
  {\gdef\sectionlabel{\thesection }}{0pt}
  {\begin{tikzpicture}[remember picture,overlay]
	\draw (1, 0) node[right] {\color{\stylecolor} \textsf{#1}};
	\fill[color=\stylecolor] (0,-0.35) rectangle (0.7, 0.35);
	\draw (0.35, 0) node {\color{white} \textsf{\sectionlabel}};
       \end{tikzpicture}
  }
\titlespacing*{\section}{0pt}{15pt}{15pt}

\begin{equation*}
\begin{tikzpicture}
\draw[very thick, color=\stylecolor] (0.0\textwidth, -5.75) -- (0.99\textwidth, -5.75);
\end{tikzpicture}
\end{equation*}

\section{Introduction}
Recently, direct evidence \cite{crossno, lucas3} has been found for the long sought Dirac fluid in graphene -- a strongly interacting quasi-relativistic  plasma of thermally excited electrons and holes \cite{vafek, schmalian}.   This Dirac fluid is of great interest for two reasons.   Firstly, it contains features of ``relativistic" quantum critical dynamics, yet is simpler than other quantum critical points.   As quantum criticality may underlie a variety of  puzzles in condensed matter physics \cite{SSBK11}, any toy experimental and/or theoretical system which may be systematically investigated is of particular interest.   Secondly, as a strongly interacting quantum system with no sharply defined quasiparticles,  the Dirac fluid is an excellent place to observe ``relativistic" hydrodynamics in a (by now) standard solid-state system.

Hydrodynamics is the universal description of how the conserved quantities of any interacting system -- generally charge, momentum and energy -- relax to global thermal equilibrium.    It is valid on long time and length scales compared to the mean free path (or thermalization length, in the absence of quasiparticles).   In the hydrodynamic limit, the microscopic degrees of freedom are not important.   The equations of motion are simply the conservation laws for conserved quantities, and as we review in Section \ref{sec:hydro}, they are readily constructed without detailed knowledge of the microscopic system of interest.   This makes hydrodynamics a particularly powerful tool for studying strongly interacting quantum systems,  where microscopic calculations are quite hard, if not essentially impossible.

The majority of theoretical study of hydrodynamics of electrons in metals focuses on the hydrodynamic regime of ultraclean Fermi liquids of electrons \cite{andreev, succiturb, tomadin, vignale, polini, levitovhydro}.  Although such samples have a large Fermi surface with long-lived quasiparticle excitations, these quasiparticles do interact weakly and so hydrodynamics ought to be valid on long time scales compared to the quasiparticle-quasiparticle collision rate.   This analogue of the hydrodynamics of classical gases (which obey the same Navier-Stokes equations as ordinary liquids) is, however, usually absent in metals  -- electron-phonon coupling, umklapp processes and impurities are all generally the dominant mechanisms for quasiparticle scattering.   More recently, in ultraclean samples, the hydrodynamics of a Fermi liquid of electrons has been experimentally observed in multiple different metals \cite{molenkamp, bandurin, mackenzie}.     

The Dirac fluid of interest in this paper is more complicated than the Fermi liquid of quasiparticles observed experimentally in metals with large Fermi surfaces \cite{molenkamp, bandurin, mackenzie}.   Firstly, like classical liquids such as water, we cannot systematically use kinetic theory to compute its properties: quasiparticles are not parametrically long-lived excitations.  Secondly, the linearized hydrodynamics of the Dirac fluid looks identical to a relativistic fluid near charge neutrality \cite{muller1, muller2, muller4, muller3}.  The resulting equations of motion and associated phenomena are distinct from usual (Galilean-invariant) fluids.     

So far, most theoretical and experimental work on the hydrodynamics of electrons in metals focuses on steady-state flows.   This limit neglects the effects of the elementary hydrodynamic excitation -- the sound mode.   This is the fundamental excitation of a clean fluid and is parametrically long lived at low energies.   Hence, it is natural to ask whether or not such sound waves could be detected in an electron fluid in a metal.   

There are other hydrodynamic systems in condensed matter which are more accessible experimentally.  Liquid $^3$He is an atomic Fermi liquid, where hydrodynamic effects due to interactions are readily observable (in contrast to the Fermi liquid of electrons in metals).   The transition between non-hydrodynamic ``zero sound" modes and ``first" sound modes (the classical sound described above) has been observed long ago in liquid $^3$He \cite{abel}.      It may also be possible to observe hydrodynamic sound in strongly interacting cold atomic gases, where other hydrodynamic behavior has been observed \cite{cao}.  

This paper is focused on the study of hydrodynamics in metals.  We present a plausible set-up for an experiment which could detect key qualitative signatures of sound modes in a charge-neutral electron-hole plasma.   In contrast to the other quantum fluids mentioned above, disorder, long-range Coulomb interactions, and electron-phonon coupling all readily spoil the propagation of pure sound modes in an electron fluid in metals.    Our aim is to quantitatively describe how these effects complicate and distort the propagation of sound waves.

\subsection{The Hydrodynamic Limit in Charge-Neutral Graphene}
As experiments on charge neutral graphene are the most likely to realize the physics proposed above, let us provide a few more details about the Dirac fluid in graphene.   The Dirac fluid consists of linearly dispersing fermions with dispersion relation $\epsilon(q) = v_{\mathrm{F}}|q|$, interacting via instantaneous long range Coulomb interactions with potential $V(r) = \alpha/r$ \cite{vafek, schmalian}.  The thermalization length scale $l_{\mathrm{th}}$ scales as \begin{equation}
l_{\mathrm{th}} \sim \frac{\hbar v_{\mathrm{F}}}{k_{\mathrm{B}}T\alpha^2} \sim 100 \; \mathrm{nm},  \label{lth}
\end{equation}
We have assumed that the effective fine structure constant\footnote{The precise value depends on the particular experimental set-up, but it is expected to be large enough that perturbative approaches formally break down.}  \begin{equation}
\alpha \sim \frac{1}{137} \frac{c}{v_{\mathrm{F}} \epsilon_{\mathrm{r}}}
\end{equation} where $c/v_{\mathrm{F}} \approx 300$ and $\epsilon_{\mathrm{r}} \sim 1$ is a dielectric constant. For details on the relevant dimensional scales in graphene, see Appendix \ref{app:dimanal}.  Up to the moderate factor of $\alpha^{-2}$, (\ref{lth}) agrees with the prediction of quantum critical theories \cite{SSBK11}.   (\ref{lth}) is a parametrically shorter thermalization length than that in a Fermi liquid, by a factor of $k_{\mathrm{B}}T/E_{\mathrm{F}}$.

In fact, $\alpha$ decreases with temperature, as Coulomb interactions are marginally irrelevant \cite{vafek, schmalian}.   However, this decrease is logarithmically slow, and hence does not lead to $\alpha \ll 1$ at room temperature.   As all first principles computations of the properties of the Dirac fluid rely on kinetic theory, they are perturbative computations in $\alpha$.   We hence are of the opinion that the best way to determine the hydrodynamic and thermodynamic properties of the Dirac fluid is to directly measure them.

The Dirac fluid is readily realized (in principle) in monolayer graphene, a two-dimensional honeycomb lattice of carbon atoms where sublattice symmetry protects multiple ``species" of Dirac fermions.   Due to ineffective screening \cite{andrei, lanzara},  these interactions are not renormalized away as effectively as in an ordinary Fermi liquid:  instead, we form the Dirac fluid.   As there is no obvious hydrodynamic experiment to detect these different species, we expect the Dirac fluid to consist of a single relativistic fluid.    

In order to detect hydrodynamics cleanly, we require that (\ref{lth}) is the shortest length scale in the problem.    The competing length scales include electron-phonon scattering lengths, as well as the density of charge puddles.    The latter is much more serious.   The Dirac fluid in graphene is subject to inhomogeneities in the charge density, commonly called charge puddles \cite{sarmachargepuddles, crommie}.   This inhomogeneous chemical potential is a direct consequence of charged impurities, primarily in the substrates on which a single layer of graphene is laid.   If the amplitude of the local chemical potential obeys $|\mu| \gg T$,  then the local physics is that of a Fermi liquid, and not the Dirac fluid.   Due to recent materials advances \cite{xue},  the size of charge puddles can now be made at least as large than (\ref{lth}), and the amplitude can be made small enough that the regime of $\mu <T$ can be reached \cite{crossno}.   In such a regime, it is natural to approximate that the Dirac fluid exists, and locally reaches thermal equilibrium at temperature $T$, with the chemical potential varying on length scales large compared to (\ref{lth}) \cite{lucas3}.    On length scales large compared to (\ref{lth}), the only degrees of freedom which are relevant are the hydrodynamic modes, including the sound mode.

Let us also briefly note that in the literature, a third ``slow" mode called the imbalance mode is often studied \cite{foster, svinstov, narozhny}.    The imbalance mode consists of the number density of electrons and holes together -- hence, combined with conservation of charge, we conclude that both electrons and holes are separately conserved.   This mode is related to the fact that accounting for two-body collisions alone it is not possible to have processes such as $\mathrm{e} \rightarrow \mathrm{e} + \mathrm{e}  + \mathrm{h}$.   This is a consequence of the conservation laws and the relativistic dispersion relation -- the ``collinear" scattering event which is allowed occupies a negligible fraction of phase space.   However, higher order processes are certainly less constrained \cite{foster}, and so this imbalance mode is not an exact conserved quantity.    Since the parameter $\alpha$ in (\ref{lth}) is not small,  higher order processes are not guaranteed to be negligible.  Hence, we do not consider this imbalance mode in our hydrodynamic description.

\subsection{Setup}
In this paper, we will develop a theory of the response of a (semi-)realistic quantum critical electron fluid, assuming approximate Lorentz invariance, to a localized finite-frequency drive.   Some qualitative features of this theory will remain true for quantum critical points with other forms of scale invariance \cite{metlitski1, metlitski2}.   In particular, some of these novel critical points are charge-neutral \cite{egmoon}, where we expect detecting sound modes to be especially clean.    Given the experimental advances described above however, we will focus on the possibility of detecting sound modes in the Dirac fluid in graphene.

\begin{figure}[t]
\centering
\includegraphics[width=4in]{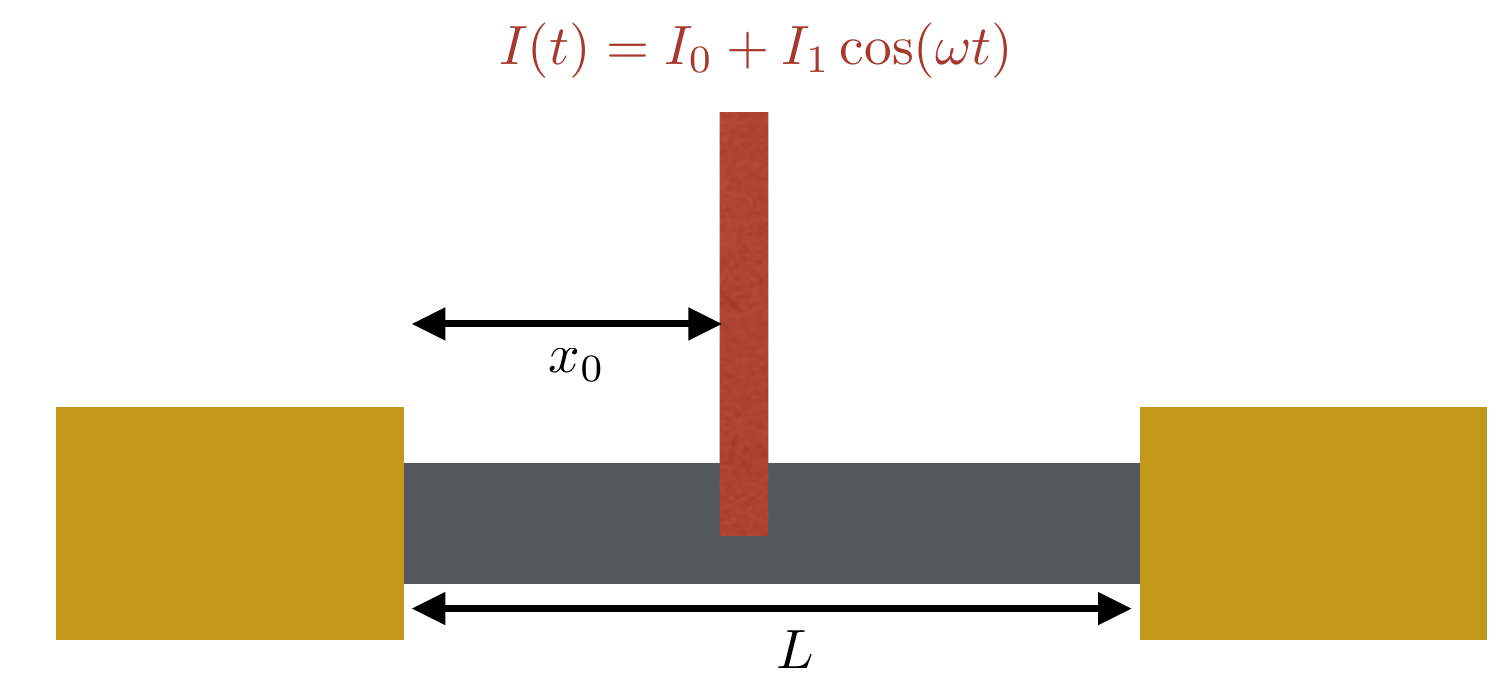}
\caption{A metallic slab of length $L$, depicted in gray, is connected to a large bath (depicted in gold).   An energy source, which we take to be a laser with time-dependent intensity $I(t)$, shines locally on the metal,  with a spot size much smaller than $L$ or $x_0$, the distance to the left contact.}
\label{fig:setup}
\end{figure}

Our proposed experimental setup is depicted in Figure \ref{fig:setup}.   A modulated source of light pumps energy into the electronic fluid at a localized region in a metal.     We then measure the electric current flowing through one-dimensional contacts \cite{wang13} at the edges of the metal.   If the laser intensity is small,  then it is reasonable to treat the electronic fluid in a linear response regime.    The electric currents measured in the contacts will consist of a dc response (which we are not interested in) and an ac response, whose magnitude will be the focus of this paper.

The signal we are after is depicted in Figure \ref{fighydroclean}.   Upon changing the modulation frequency $\omega$ of the energy source, the electron-hole plasma in a clean sample where disorder and electron-phonon effects are not substantial will be driven at a resonant frequency,  leading to a large jump in the magnitude of the current flowing out of the edges.    This resonance is directly associated with the normal modes of electronic classical sound waves, and is the electronic analogue to the resonances that occur when air is excited in long pipes such as musical instruments.   Such resonances cannot be seen if the electronic dynamics is dominated by diffusive ``Ohmic" processes.

We also note that similar experimental setups (at $\omega=0$) have been performed to uncover ``hot carrier" dynamics in graphene \cite{song2, ma1, song, johannsen, ma2}.   In this limit the electrons behave diffusively, albeit with some unexplained signals reported near charge neutrality \cite{ma2}.   In our simpler set-up, however, there is very little signal of interest at $\omega=0$.   

We will briefly comment in the conclusion on alternative measurements which may be simpler, but the key physics of interest is easiest to understand in this simple set-up.

\subsection{Main Results}
We now state the main results of this paper.   While basic results which are similar have appeared before in the literature \cite{levitovsound}, our hydrodynamic framework differs subtly, as we will explain later.   We will review the correct hydrodynamic framework for the Dirac fluid in Section \ref{sec:hydro}.
\begin{itemize}
\item Working under the assumption that the electronic dynamics is collective (no single-particle effects),  the observation of resonances in the finite frequency response is a smoking gun signature for the hydrodynamic sound mode: see Section \ref{sec:homo}.   Assuming the absence of charge puddles, we can solve the hydrodynamic equations exactly: see (\ref{eq:jmain}) and Appendix \ref{app:anal}.   Assuming that dissipation is weak enough to observe the $n^{\mathrm{th}}$ sound mode, a resonance will be measured at frequency\begin{equation}
\frac{\omega_n}{2\mpi} = \frac{n\mpi v_{\mathrm{F}}}{2\mpi \sqrt{d} L} \approx n \times  \frac{1\; \mmu \mathrm{m}}{L}  \times 0.3\; \mathrm{THz}.  \label{eq:resfreqs}
\end{equation}
The numbers presented are estimates for graphene, where the number of spatial dimensions $d=2$.  A realistic sample may have $L=10$ $\mmu$m, in which case observing the first sound mode requires reading off the electric current flowing across the contacts at 30 GHz.   
\item Experimental observation of acoustic resonances gives semi-quantitative upper bounds on the viscosity $\eta$ of the electron fluid: see (\ref{etas}).
\item Taking into account long range Coulomb interactions, the sound modes morph into plasmonic modes with different dispersion relations.   The resonances in (\ref{eq:resfreqs}) will shift to new frequencies.   Such modes have been found before in theoretical models \cite{levitovsound}, though it appears that our formula for the lifetime of these plasmons is new. 
\item Although momentum loss due to phonons and disorder may appear similar in the $\omega \rightarrow 0$ limit,  at finite $\omega$ the response of the fluid becomes qualitatively different depending on the mechanism for momentum relaxation.  If the dominant source of momentum relaxation is long-wavelength charge density inhomogeneity, and not coupling to acoustic phonons, the observed sound resonances can be parametrically larger than predicted using simpler models, making experimental detection much easier:  see Section \ref{sec:dis}.
\item From a theoretical perspective, the classical Anderson localization of sound waves in inhomogeneous electron fluids is rather interesting, with the localization length a non-monotonic function of the frequency.    Localization can lead to a complete breakdown of momentum relaxation time approximations, which are commonly employed to mimic the interplay of the electron fluid with disorder:  see Section \ref{sec:dis1}.   The breakdown of this approximation occurs at at a frequency scale given in (\ref{omegafail}), which can be arbitrarily small.
\item Depending on the equations of state of the Dirac fluid, and the quality of a graphene sample, it may be possible to quantitatively extract the speed of sound through our experimental setup.   The speed of sound in the Dirac fluid is given by (\ref{eq:vs}) and contains no fitting parameters.   Since (\ref{eq:vs}) is an irrational multiple of $v_{\mathrm{F}}$,  it would be unambiguous to experimentally distinguish between single-particle resonances, observed in carbon nanotubes in \cite{mceuen}, and hydrodynamic sound resonances.
\end{itemize}

In this paper, we will generically work in units where $\hbar = v_{\mathrm{F}} = k_{\mathrm{B}} = e = 1$.   When presenting numerical results, we will further work in units where the background fluid temperature is set to $T=1$ -- this fixes all quantities to be dimensionless.   Hydrodynamics is applicable when $\omega/T \ll 1$.   The units may always be restored straightforwardly with dimensional analysis -- see Appendix \ref{app:dimanal}.   We will discuss what we believe are the major experimental challenges  in the conluding section.   Technical details of computations are placed in appendices.

\section{Hydrodynamic Linear Response Theory}\label{sec:hydro}
In this section, we review the hydrodynamics of quantum critical fluids with emergent Lorentz invariance \cite{hkms}.   In linear response, this hydrodynamics describes the electron-hole plasma in charge-neutral graphene \cite{muller4, muller2}.   These equations are the conservation of charge, along with the conservation of energy and momentum up to external sourcing (including that due to the external chemical potential inhomogeneity): \begin{subequations}\label{eq:hydromain}\begin{align}
\partial_t T^{tt} + \partial_i T^{ti} &= J^i\partial_i\mu_0 + \mathcal{S}, \\
\partial_t T^{ti} + \partial_j T^{ij} &= J^t \partial_i \mu_0 - \gamma v^i, \\
\partial_t J^t + \partial_i J^i &= 0,
\end{align}\end{subequations}
where $T^{\mu\nu}$ is the stress-energy tensor,  $J^\mu$ is the charge current, $\mu_0$ is an externally imposed chemical potential.   $\mathcal{S}$ is an external source of energy -- in the setup described in Figure \ref{fig:setup}, this is due to a laser causing local heating of the electron-hole plasma.   The parameter $\gamma>0$ accounts for the loss of momentum due to electron-phonon coupling.    We will keep the number of spatial dimensions $d$ of the fluid generic for much of the paper;  obviously, for the application to graphene, one can set $d=2$.

For simplicity  in this paper, we will assume that the disorder is only inhomogeneous along one spatial direction, which we denote as $x$.   This is for computational simplicity, although such an assumption is likely appropriate for highly oblong samples of graphene in experiment.    Hence, the disorder profile is a simple function $\mu_0(x)$.

Let us begin by finding a static solution to (\ref{eq:hydromain}).   The equations of state of relativistic hydrodynamics in the Landau frame are: \begin{subequations}\label{eq:hydrost}\begin{align}
T^{tt} &= \epsilon + \mathrm{O}\left(v^2\right), \\
T^{tx} &= (\epsilon+P) v + \mathrm{O}\left(v^3\right), \\
T^{xx} &= P - \eta^\prime \partial_x v + \mathrm{O}\left(v^2\right), \\
J^t &= n+ \mathrm{O}\left(v^2\right), \\
J^x &= n v - \sigma_{\textsc{q}}\left(\partial_x  (\mu-\mu_0) - \frac{\mu}{T} \partial_x  T\right) + \mathrm{O}\left(v^2\right)
\end{align}\end{subequations}
with $T$ the temperature, $\mu$ the chemical potential,  $v$ the fluid velocity, $n$ the charge density, $\epsilon$ the energy density,  and $P$ the pressure, all locally defined using thermodynamics, and \begin{equation}
\eta^\prime = \zeta + \eta\left(2-\frac{2}{d}\right).
\end{equation}
with $\zeta$ and $\eta$ the bulk and shear viscosity, respectively.  Upon using the thermodynamic relation \begin{equation}
\mathrm{d}P = n\mathrm{d}\mu + s\mathrm{d}T  \label{eq:Pthermo}
\end{equation}
with $s$ the entropy density, we see that (\ref{eq:hydromain}) is solved by  $T=T_0$,  $\mu = \mu_0(\mathbf{x})$, and $v=0$ \cite{lucas, lucas3}.    We will find the identity \begin{equation}
\epsilon = dP \label{eq:edp}
\end{equation}useful,  and will often write $\epsilon+P = (d+1)P$ interchangably.

In linear response about this solution, we define $\tilde \mu = \mu-\mu_0$,  $\tilde T= T-T_0$, and $\tilde v = v_x$;  indeed, tilded variables henceforth denote first order quantities in linear response.   The linearized charge and energy-momentum currents are: \begin{subequations}\label{eq:hydrost}\begin{align}
\tilde T^{tt} &= \frac{\partial \epsilon}{\partial \mu} \tilde \mu + \frac{\partial \epsilon}{\partial T} \tilde T = dn\tilde \mu + ds\tilde T, \\
\tilde T^{tx} &= (\epsilon+P)\tilde v = (d+1)P\tilde v, \\
\tilde T^{xx} &= \tilde P - \eta^\prime \partial_x \tilde v, \\
\tilde J^t &= \frac{\partial n}{\partial \mu} \tilde \mu + \frac{\partial n}{\partial T} \tilde T, \\
\tilde J^x &= n\tilde v - \sigma_{\textsc{q}}\left(\partial_x \tilde \mu - \frac{\mu}{T} \partial_x \tilde T\right)
\end{align}\end{subequations}
For simplicity, we have also dropped the 0 subscript on the background $\mu$, and will do so for most of this paper.   Note that \begin{equation}
\partial_x \tilde P = n\partial_x \tilde \mu + \tilde n \partial_x \mu + s\partial_x \tilde T.
\end{equation}

As described in the introduction, we assume in our set-up that $\mathcal{S}$ is periodically driven in time at angular frequency $\omega$.   Using standard tricks, we hence solve the linear response equations using the complex-valued source \begin{equation}
\mathcal{S} = \mathcal{S}_0(x) \mathrm{e}^{-\mathrm{i}\omega t},
\end{equation}
and look for solutions where $\tilde \mu(x,t) = \tilde \mu(x)\mathrm{e}^{-\mathrm{i}\omega t}$, etc. Hence, we may replace $\partial_t \rightarrow -\mathrm{i}\omega$ in (\ref{eq:hydromain}).    The real part of the solutions are physical, but we will often study the complex modulus of the response, which corresponds to the overall amplitude of temporal oscillations.    When doing analytic calculations, we will consider the following simplified model for the local injection of energy: \begin{equation}
\mathcal{S}_0(x) = \Gamma \; \mdelta(x-x_0),
\end{equation}
but more generally we can also solve these equations with a smoothed out $\mdelta$ function, and will indeed do so in our numerics.

Putting it all together, we need to solve the elliptic linear ordinary  differential equations \begin{subequations}\label{eq:linear}\begin{align}
\partial_x \left((\epsilon+P)\tilde v\right) &= \left[n\tilde v - \sigma_{\textsc{q}}\left(\partial_x \tilde \mu - \frac{\mu}{T} \partial_x \tilde T\right)\right]\partial_x \mu  + \mathcal{S}_0(x) + \mathrm{i}\omega d\left( n \tilde \mu + s \tilde T \right), \\
 n \partial_x \tilde \mu + s\partial_x \tilde T - \partial_x \left(\eta^\prime \partial_x \tilde v\right) &= \left[\mathrm{i}\omega (\epsilon+P) - \gamma \right]\tilde v, \label{eq:linearb} \\
\partial_x\left[n\tilde v - \sigma_{\textsc{q}}\left(\partial_x \tilde \mu - \frac{\mu}{T} \partial_x \tilde T\right)\right] &= \mathrm{i}\omega \left(\frac{\partial n}{\partial \mu} \tilde \mu + \frac{\partial n}{\partial T} \tilde T\right) 
\end{align}\end{subequations} in a finite domain, which we take to be $0\le x \le L$. We assume the boundary conditions \begin{equation}
\tilde \mu(x=0,L) = \tilde T(x=0,L) = 0,  \label{bc1}
\end{equation}
corresponding to the physical assumption  that near the contacts, the electron fluid is well thermalized to the external bath.    These boundary conditions are sufficient for the hydrodynamic equations defined previously to be well-posed, as we explain in Appendix \ref{app:bc}.    We solve these equations using standard numerical techniques, which we review in Appendix \ref{app:num}.

It is helpful to keep in mind the following simple symmetries of our hydrodynamic equations (below $\lambda>0$ is a constant re-scaling parameter):  \begin{subequations}\label{eq:rescale}\begin{align}
&P \rightarrow \lambda P, \;\;\;\;\; \sigma_{\textsc{q}} \rightarrow \lambda \sigma_{\textsc{q}}, \;\;\;\;\; \eta^\prime \rightarrow \lambda \eta^\prime  , \;\;\;\;\; \gamma \rightarrow \lambda \gamma \\
&  x\rightarrow \lambda x \;\;\;\;\; \omega \rightarrow \lambda^{-1}\omega, \;\;\;\;\;  \gamma \rightarrow \lambda^{-1}\gamma, \;\;\;\;\;  \sigma_{\textsc{q}} \rightarrow \lambda \sigma_{\textsc{q}}, \;\;\;\;\; \eta \rightarrow \lambda \eta.
\end{align}\end{subequations}
The latter is especially helpful, as it means that the effects of dissipation can be (in principle) made arbitrarily large or small, locally (neglecting $\gamma$, as we will often do).

When doing numerics, we must be more specific about the equations of state.   We will follow \cite{lucas3} and employ the simple equations of state \begin{subequations}\begin{align}
\frac{\epsilon}{d} = P &= \frac{C_0}{d+1} T^{d+1} + \frac{C_2}{2}\mu^2 T^{d-1}, \\
n &= C_2 \mu T^{d-1}, \\
s &= C_0 T^d + (d-1)\frac{C_2}{2} \mu^2 T^{d-2}, \\
\eta^\prime &= \eta_0 T^d, \\
\sigma_{\textsc{q}} &= \sigma_0 T^{d-2}.
\end{align}\end{subequations}

\section{Response of a Homogeneous Fluid}\label{sec:homo}
We now exactly solve these equations under the assumption that the background chemical potential is constant. 

\subsection{Normal Modes}\label{sec:normalmode}
When the background fluid is translation invariant, it is natural to study hydrodynamics by looking for normal modes -- solutions to the equations of motion with $x$ and $t$ dependence given by  $\mathrm{e}^{\mathrm{i}qx-\mathrm{i}\omega t}$.   Even in our finite size, locally driven system (which breaks translation invariance), knowledge of such normal modes is useful.  We thus begin with a thorough review of the theory of such normal modes in a relativistic fluid such as the Dirac fluid.

Let us choose the variables $\tilde P$, $\tilde n$ and $\tilde v$ as our dynamical hydrodynamic variables.   $\tilde P$ and $\tilde n$ may be related to $\tilde \mu$ and $\tilde T$ using (\ref{eq:hydrost}) and (\ref{eq:Pthermo}).    (\ref{eq:linear})  reduces to an algebraic set of equations \begin{subequations}\label{eq:sound}\begin{align}
0 &= -\mathrm{i}\omega \tilde n + \mathrm{i}q n\tilde v +q^2D \tilde n - q^2 C\tilde P, \\
0 &= -\mathrm{i}\omega d \tilde P + (d+1)\mathrm{i}qP\tilde v, \label{eq:s4b} \\
0 &= [\gamma-\mathrm{i}\omega(d+1)P]\tilde v + \mathrm{i}q\tilde P + \eta q^2 \tilde v.
\end{align}\end{subequations} 
where we have defined \begin{subequations}\begin{align}
C &= \frac{\sigma_{\textsc{q}}}{s\partial_\mu n - n\partial_Tn} \left(\frac{\partial n}{\partial T}+ \frac{\mu}{T}\frac{\partial n}{\partial \mu}\right), \\
D &= \frac{\sigma_{\textsc{q}}}{s\partial_\mu n - n\partial_Tn} \frac{(d+1)P}{T}.  \label{eq:Dcons}
\end{align}\end{subequations}
The equations governing $\tilde P$ and $\tilde v$ decouple from the others.     Note that the thermodynamic equations are constrained so that $D>0$ \cite{kovtun}.

Studying the solution to (\ref{eq:sound}) assuming non-vanishing $\tilde P$ and $\tilde v$, we obtain dissipative sound waves as $\omega,q\rightarrow 0$, with the dispersion relation \begin{equation}
\omega = \pm\sqrt{\frac{q^2}{d} - \left(\frac{\gamma + \eta^\prime q^2}{2(d+1)P}\right)^2} - \mathrm{i}\frac{\gamma + \eta^\prime q^2}{2(d+1)P}
\end{equation}
In the limit where $\gamma \ll P\omega$, we obtain ordinary sound waves damped by viscosity: \begin{equation}
\omega \approx  \pm v_{\mathrm{s}} |q| - \frac{\mathrm{i}\eta^\prime q^2}{2(d+1)P} + \mathrm{O}\left(q^3\right),
\end{equation}
with the speed of sound given by \begin{equation}
v_{\mathrm{s}} = \frac{1}{\sqrt{d}}= \frac{v_{\mathrm{F}}}{\sqrt{d}}.  \label{eq:vs}
\end{equation}
This is a robust prediction of relativistic hydrodynamics, assuming the absence of dimensionful scales other than $\mu$ and $T$, and follows from (\ref{eq:edp}).   When $\gamma \gg P\omega$, we instead obtain a diffusive mode \begin{equation}
\omega = -\mathrm{i}\left[\frac{(d+1)P}{d\gamma} - \eta^\prime\right]q^2  \label{eq:omegadiff}
\end{equation}
along with a finite lifetime mode $\omega = -\mathrm{i}\gamma/(d+1)P$.  We have implicitly assumed that $\gamma$ was small when writing down (\ref{eq:linear}), so the diffusion constant above is positive.   Similar sound waves were described in \cite{svinstov, levitovsound}, but these papers assume a different dissipative structure in hydrodynamics.   It is necessary to follow \cite{hkms} to obtain a non-vanishing electric current, under any circumstances, at charge neutrality.

In a sound wave $\tilde n$ is slave to $\tilde P$ and $\tilde v$.   There is also a diffusive charge mode with \begin{equation}
\omega = -\mathrm{i}Dq^2,
\end{equation}
in which only $\tilde n$ is non-vanishing.   

The qualitative form of the hydrodynamic modes listed above is very similar to those in Galilean-invariant fluids, including liquid $^3$He.   We review this theory in Appendix \ref{sec:appgal}.   However, the theory of Galilean-invariant fluids becomes singular if the fluid is charge-neutral -- there is no charge current or momentum density.  As a consequence, it is important to use the relativistic theory of hydrodynamics to study charge-neutral electron-hole plasma, as can be found in graphene.  This is especially important once we account for disorder in Section \ref{sec:dis}.

\subsection{Observing Resonances Near Charge Neutrality}
The set-up of interest in this paper may readily detect the sound modes described above.   To show this, we must solve the boundary value problem described in the previous section.  As this uses relatively standard techniques, we present the calculation in Appendix \ref{app:anal}.     The current at $x=0$ is given by \begin{equation}
|\tilde J| = \frac{\Gamma}{(d+1)P} \left| \frac{\mathrm{i}\omega}{\mathrm{i}\omega - Dq^2_{\mathrm{s}}} \left(n+\frac{\mathrm{i}(d+1)PCq_{\mathrm{s}}^2}{d\omega}\right) \right|\left| \frac{\sin(q_{\mathrm{s}}(L-x_0))}{\sin (q_{\mathrm{s}} L)} - \frac{\sin(q_{\mathrm{d}}(L-x_0))}{\sin (q_{\mathrm{d}} L)} \right|,  \label{eq:jmain}
\end{equation}
where we have defined
\begin{subequations}\begin{align}
\displaystyle q_{\mathrm{s}} &\equiv \omega \sqrt{\dfrac{\displaystyle 1+ \frac{\mathrm{i}\gamma}{\omega (d+1)P}}{\displaystyle\frac{1}{d} - \frac{\mathrm{i}\eta \omega}{(d+1)P}}},  \label{eq:veta} \\
q_{\mathrm{d}} &\equiv \sqrt{\frac{\mathrm{i}\omega}{D}}.  \label{eq:qD}
\end{align}\end{subequations}

\begin{figure}[t]
\centering
\includegraphics[width=4in]{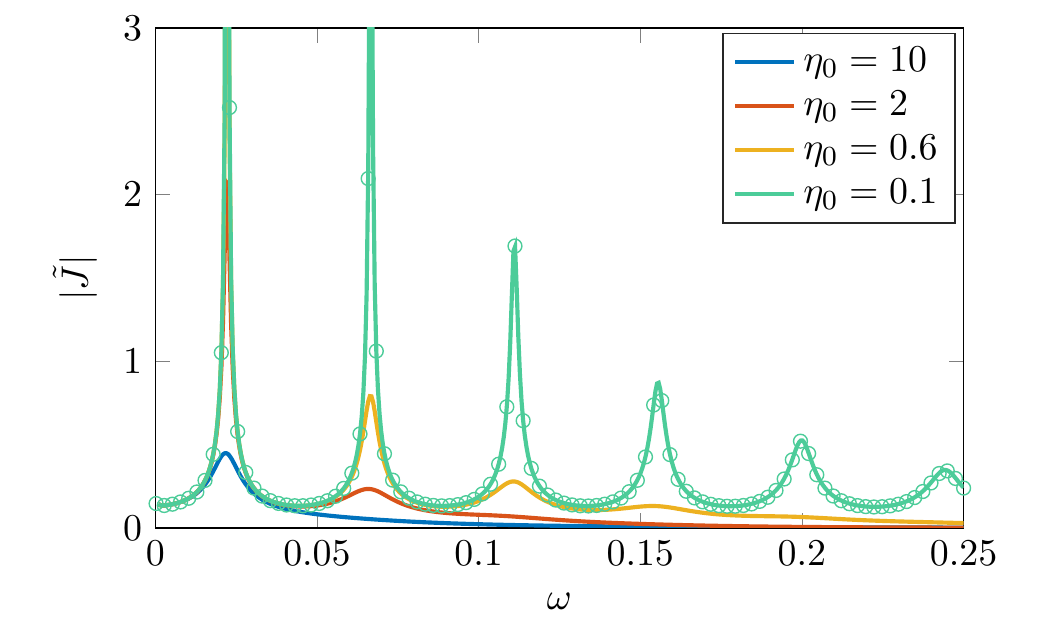}
\caption{A plot of $|\tilde J|$ as given in (\ref{eq:jmain}), assuming  $\Gamma=1$, $L=100$, $C_0=C_2=1$, $\sigma_0=0.1$, $\mu=0.3$, and $\gamma=0$, for different values of $\eta_0$.   Note how both the sharpness of the acoustic resonances and the number of observable resonances increases as the viscosity $\eta_0$ decreases.  The circles compare our numerical methods to the analytic prediction for one such value of $\eta_0$, confirming the accuracy of our numerical methods.}
\label{fighydroclean}
\end{figure}

For many purposes, it is acceptable to neglect the $q_{\mathrm{d}}$-dependent term in (\ref{eq:jmain}).  Let us temporarily make this assumption.  In the limit of small dissipation,  the response of the fluid becomes parametrically sharp as $\omega$ is tuned to a resonant frequency, which will occur when $\sin(q_{\mathrm{s}} L) \approx 0$.   Expanding $q_{\mathrm{s}}$ to linear order in $\omega$, this occurs when \begin{equation}
0 \approx \sin(\omega L\sqrt{d}) \cos \left(\frac{\mathrm{i}d^{3/2}}{2(d+1)} \frac{\eta\omega^2L}{P} + \frac{\mathrm{i}\sqrt{d}L\gamma }{2(d+1)P}\right) + \cos(\omega L\sqrt{d})\sin \left(\frac{\mathrm{i}d^{3/2}}{2(d+1)} \frac{\eta\omega^2L}{P}+ \frac{\mathrm{i}\sqrt{d}L\gamma}{2(d+1)P}\right). 
\end{equation}
Hence, if \begin{equation}
\omega = \frac{n\mpi}{L\sqrt{d}} = \frac{n\mpi v_{\mathrm{s}}}{L},
\end{equation}
we see that as $\omega \rightarrow 0$, \begin{equation}
\sin (q_{\mathrm{s}}L)= \sinh\left(\frac{d^{3/2} n^2 \mpi^2 }{2(d+1)} \frac{\eta}{PL} + \frac{\sqrt{d}L\gamma}{2(d+1)P}\right)  \label{smallsinh}
\end{equation}
And if $\tau \rightarrow \infty$ and $\eta\rightarrow 0$,  so that dissipative effects are very small,  we see  that the current $\tilde J$ flowing through the contacts, given in  (\ref{eq:jmain}), is proportional to the inverse of a small parameter.  Indeed, these resonances correspond to the normal modes of a one dimensional wave equation with standard (closed-closed) boundary conditions.   When the argument of the sinh becomes large, then we cannot observe any more sharp resonances.     We numerically plot (\ref{eq:jmain}) for various values of $\eta_0$ in Figure \ref{fighydroclean}.

As the assumption that Coulomb interactions are negligible is best near the charge neutrality point (as we will discuss later), we now assume that $\mu=0$ and discuss the practical consequences of (\ref{smallsinh}) in $d=2$ (relevant for experiments on graphene).   Let us first assume that the sample is perfectly clean.  At charge neutrality,  then $(d+1)P=Ts$, and sharp resonances occur when \begin{equation}
\frac{\eta}{s} \lesssim \frac{TL}{10n^2} \sim \frac{1}{10n^2} \frac{L}{l_{\mathrm{th}}} .  \label{etas}
\end{equation}
Ideally, we could observe at least $n=3$ resonances, meaning that experimentally $L> l_{\mathrm{th}} \times 100 \eta/s$.   If we use the experimentally-derived estimate that $\eta/s \sim 10$ \cite{lucas3},  (\ref{lth}) implies that $L>100$ $\mmu$m, which is an order of magnitude too large for many experiments.   Of course, if we only wish to see the first resonance, or (as postulated in \cite{lucas3}) $\eta/s$ is in reality closer to 2,  then $L>10$ $\mmu$m may be sufficient, and within experimental reach.
Alternatively, we may say that if $n$ resonances are observed, then the ratio $\eta/s$ is heuristically bounded from above by (\ref{etas}).

\begin{figure}[t]
\centering
\includegraphics[width=3.2in]{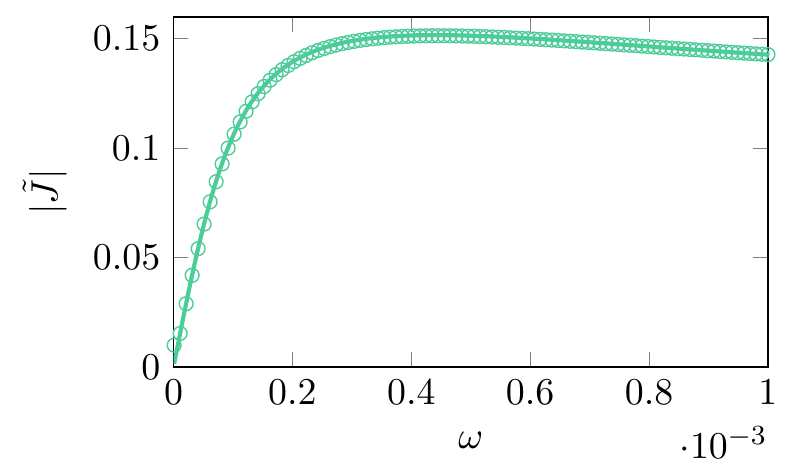}
\caption{A plot of $|\tilde J|$ as given in (\ref{eq:jmain}), and a comparison to numerics, for $\eta_0=0.1$, and otherwise the same parameters in Figure \ref{fighydroclean}.  Here we focus on the limit $\omega \rightarrow 0$, where we see that $\tilde J\rightarrow 0$.}
\label{fighydroclean3}
\end{figure}

Finally, let us also note that in the limit $\omega \rightarrow 0$,  $\tilde J$ as given in (\ref{eq:jmain}) vanishes -- see Figure \ref{fighydroclean3}.   Hence, the signal we are looking for is strictly finite frequency.   However, the frequency scale at which the vanishing of this signal disappears is $\omega \sim D/L^2$, which (for realistic parameters for graphene) is about 1 GHz.  Hence, for all practical purposes, this effect is incredibly suppressed by the time we reach the scale of sound resonances at $\sim$ 30 GHz.

\subsection{The Diffusive Limit}

\begin{figure}[t]
\centering
\includegraphics[width=4in]{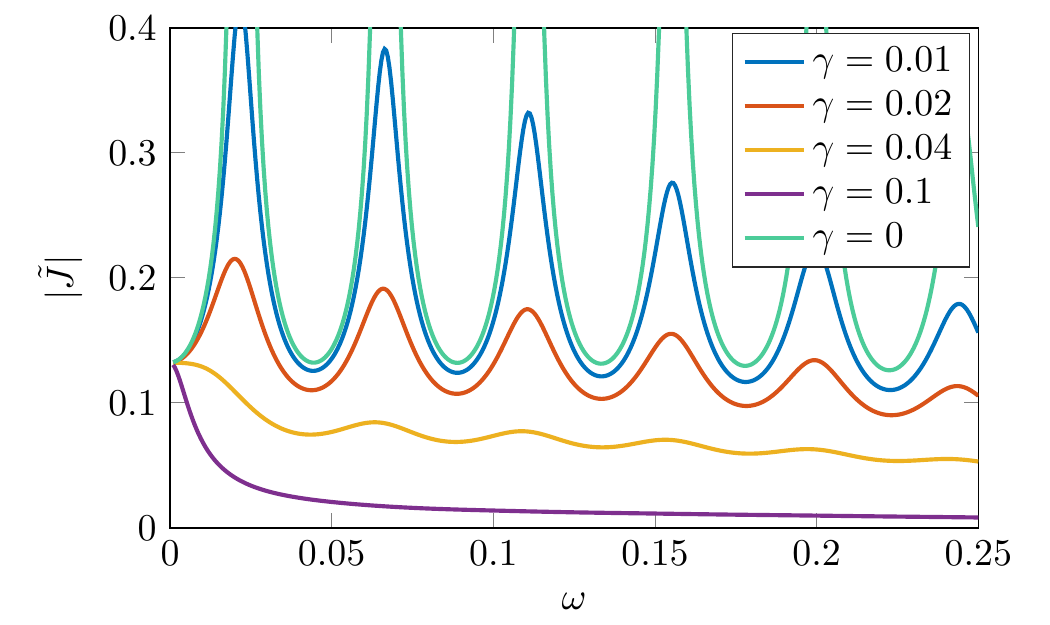}
\caption{A plot of $|\tilde J|$ as given in (\ref{eq:jmain}), assuming $\Gamma=1$, $L=100$, $C_0=C_2=1$, $\sigma_0=0.1$, $\mu=0.3$, and $\eta_0=0.1$, for different values of $\gamma$.    Upon increasing $\gamma$, the strength of the observed signal decreases exponentially, and fewer resonances are easily visible.   Furthermore, the maximal strength of the observable signal is set by $\gamma$.}
\label{fighydrogamma}
\end{figure}

Figure \ref{fighydrogamma} shows the consequences of adding $\gamma>0$.   As expected, the presence of electron-phonon coupling degrades the sharpness of the acoustic resonances of the electron fluid, surprsingly quickly, even when $\gamma \sim 0.04$.  We can estimate $\gamma$ in charge-neutral graphene, following \cite{hwang07, viljas}, assuming that the electrons are non-interacting: \begin{equation}
\frac{\gamma}{(d+1)P} \sim \frac{\mathcal{D}^2 (k_{\mathrm{B}}T)^2}{\hbar^3 \rho_{\mathrm{m}} v_{\mathrm{F}}^2 v_{\mathrm{ph}}^2} \sim 10^{10} \; \mathrm{s}^{-1} \sim 10^{-3} \; (\text{natural units}).
\end{equation}
Here $\mathcal{D}$ is the deformation potential of graphene, $\rho_{\mathrm{m}}$ is the mass density of the graphene crystal, and $v_{\mathrm{ph}} \sim 0.02 v_{\mathrm{F}}$ is the speed of acoustic phonons; appropriate numbers are given in \cite{hwang07}.   This simple estimate suggests that the effects of $\gamma$ are rather negligible for a realistic experiment.   

If $\gamma \gtrsim (d+1)P \omega$, then we never see any resonances.  
Defining \begin{equation}
\tilde D = \frac{(d+1)P}{d\gamma},
\end{equation}
and
\begin{equation}
q_\omega \equiv \sqrt{-\frac{\mathrm{i}\omega}{\tilde D}},
\end{equation}
we obtain from (\ref{eq:jmain}) that the current flow at the endpoints is approximately given by\begin{equation}
|\tilde J | \sim \left|\frac{\sinh(q_\omega (L-x_0))}{\sinh(q_\omega L)}\right|.
\end{equation}
It is straightforward to check that this function of $\omega$ is monotonically decreasing, and will admit no resonances.   

At higher frequencies, neglecting viscosity, we may expand $q_{\mathrm{s}}$  to first non-trivial order in $\omega$.   (\ref{eq:jmain}) and (\ref{smallsinh}) lead to \begin{equation}
|\tilde J| \sim \exp\left[-\frac{\sqrt{d}x_0\gamma}{2(d+1)P}\right],  \label{eq:expsupgamma}
\end{equation}
and so even at high frequencies,  the remnants of the homogeneous (electron-phonon) momentum relaxation channel leads to exponential suppression of the signal.   This effect is readily observed in Figure \ref{fighydrogamma}.   We will see that the inhomogeneous momentum relaxation channels do \emph{not} have this property.

The absence of resonances in a diffusive limit implies that the detection of non-monotonic behavior in $\tilde J(\omega)$ is a powerful test for clean hydrodynamics in an electronic fluid.   In contrast, Ohmic models such as ``hot carrier dynamics" \cite{song2} predict monotonically decreasing $\tilde J(\omega)$ once $\omega \gg D/L^2$.

\section{Coulomb Interactions}
In this section, we discuss the role of long-range Coulomb interactions more carefully.   As we will see,  unlike for $\omega=0$ phenomena, at finite $\omega$ it is important to account for such interactions.   As a ``worst case scenario", we will assume that long-range Coulomb interactions are completely unscreened.  In reality, thermal fluctuations can lead to additional screening, analogous to the Debye screening of ions in water \cite{sarma2009}, on the longest wavelengths.   Secondly,  in an experiment such as \cite{crossno}, the presence of gates near the sample leads to image charges $\sim 600$ nm from the sample,  which modifies further the Coulomb kernel beyond this length scale.   Both of these effects will reduce the complications of Coulomb screening -- and possibly eliminate them (qualitatively) when looking for sound modes with wavelengths $\sim 10$ $\mmu$m.    Nonetheless, it is instructive to understand possible complications such Coulomb effects could lead to.

Following \cite{lucas3, muller1}, we account for long-range Coulomb interactions by replacing $\mu_0(\mathbf{x})$, the externally imposed chemical potential,  with the ``external" electrochemical potential \begin{equation}
\mu_0 - \varphi = \mu_0 - \int \mathrm{d}^d\mathbf{y} \; K(\mathbf{x};\mathbf{y}) n(\mathbf{y}),  \label{eq:defcou}
\end{equation}
where $K$ is a long-range Coulomb kernel which we will describe in more detail later.   This causes two changes.   Firstly, the background is no longer given by $\mu = \mu_0$,  but by a more complicated solution where $\mu$ varies so that $\mu(\mathbf{x}) = \mu_0 - \varphi[\mu_0]$.   Since the hydrodynamic equations depend either on $\mu$ (through the local equations of state),  or $\mu_0 - \varphi = \mu$,  it is acceptable to ``ignore" this effect through a redefinition of $\mu_0$.   Physically, this is the statement that the electronic fluid only is sensitive to the electrochemical potential.   The second change is non-trivial, however.   The linear response equations now read \begin{subequations}\begin{align}
 \left[n\tilde v - \sigma_{\textsc{q}}\left(\partial_x \tilde \mu + \partial_x (K\otimes \tilde n) - \frac{\mu}{T} \partial_x \tilde T\right)\right]\partial_x \mu   + \mathcal{S}_0(x) + \mathrm{i}\omega d\left( n \tilde \mu + s \tilde T \right) &= \partial_x \left((\epsilon+P)\tilde v\right) , \\
 n \partial_x \tilde \mu + n \partial_x (K\otimes \tilde n) + s\partial_x \tilde T - \partial_x \left(\eta^\prime \partial_x \tilde v\right) &= \left[\mathrm{i}\omega (\epsilon+P) - \gamma \right]\tilde v,  \\
\partial_x\left[n\tilde v - \sigma_{\textsc{q}}\left(\partial_x \tilde \mu + \partial_x (K\otimes \tilde n) - \frac{\mu}{T} \partial_x \tilde T\right)\right] &= \mathrm{i}\omega \left(\frac{\partial n}{\partial \mu} \tilde \mu + \frac{\partial n}{\partial T} \tilde T\right) 
\end{align}\end{subequations}
where $K\otimes \tilde n$ denotes the convolution in (\ref{eq:defcou}), but only over the linear response perturbation to the local charge density, $\tilde n = (\partial_\mu n) \tilde \mu + (\partial_ T n) \tilde T$.

We now describe the form of $K(x,y)$.  The Poisson equation governing the long range Coulomb interactions in the physical three spatial dimensions reads \begin{equation}
\left(\partial_x^2  +\partial_z^2 \right) \varphi = -4\mpi \alpha n \; \mdelta(z),  \label{eq:poisson}
\end{equation}
where $z$ is the out-of-plane direction,  we have placed the graphene sheet at $z=0$, and $\alpha$ is the effective fine structure constant, discussed in the introduction.  In the infinite plane, one finds \begin{equation}
K(x,y) = -2\alpha \log |x-y|.  \label{eq:Kxy}
\end{equation} 
It is helpful to Fourier transform this expression: \begin{equation}
(K\otimes \tilde n)(q) = \frac{2\mpi\alpha}{|q|}\tilde n(q).  \label{eq:2paq}
\end{equation}
In a finite domain, subject to the boundary conditions $\tilde\varphi(x=0,L)=0$, $K(x,y)$ can be expressed as a Fourier series, which we explicitly compute in Appendix \ref{sec:appcou}.   For distances $|x-y| \ll L$, it reduces approximately to (\ref{eq:Kxy}).

\subsection{Sound Waves and Plasmons}
The normal modes of a fluid, accounting for long range Coulomb interactions, may be found analogously to Section \ref{sec:normalmode}.   We state the results -- now explicitly plugging in for $d=2$.   Assuming that $n\ne 0$, the frequencies of the normal modes are given by the solutions to the equation \begin{equation}
0=\gamma - 3\mathrm{i}\omega P + \eta^\prime q^2 + \frac{3\mathrm{i}Pq^2}{2\omega } + \frac{\mathrm{i}\alpha n|q|}{2\omega}\frac{2\mathrm{i}n\omega - 3CPq^2}{\mathrm{i}\omega - Dq^2 - \sigma_{\textsc{q}}\alpha|q|}
\end{equation}
We find a diffusive mode given by \begin{equation}
\omega = -\mathrm{i} \frac{9\sigma_{\textsc{q}} P^2 \partial_\mu n }{2T(n^2+\sigma_{\textsc{q}}\gamma)(s\partial_\mu n - n\partial_\mu s)}  q^2 + \cdots  \label{eq:coudiff}
\end{equation}
and propagating plasmon modes with \begin{equation}
\omega \approx \frac{-\mathrm{i}[\gamma + 6\mpi \alpha P\sigma_{\textsc{q}}|q|] \pm \sqrt{24\mpi \alpha Pn^2|q|-[\gamma + 6\mpi \alpha P\sigma_{\textsc{q}}|q|] ^2}}{6P} + \mathrm{O}\left(q^{3/2},\eta^\prime\right).
\end{equation}
When $\gamma \ll 6\mpi \alpha P\sigma_{\textsc{q}}|q|$: \begin{equation}
\omega = \sqrt{\frac{2\mpi \alpha n^2}{3P}|q|}   - \mathrm{i}\mpi \alpha \sigma_{\textsc{q}} |q| + \cdots,
\end{equation}
and when $\gamma \gg 6\mpi \alpha P\sigma_{\textsc{q}}|q|$, we find a damped mode with $\omega = -\mathrm{i}\gamma/3P$, and a ``diffusive" mode with \begin{equation}
\omega = -\mathrm{i} \frac{2\mpi \alpha n^2}{\gamma} |q|
\end{equation}
In the dissipationless limit where $\gamma=\eta^\prime=\sigma_{\textsc{q}}=0$, the propagating modes have dispersion relation \begin{equation}
\omega^2 = \frac{q^2}{2} + \frac{2\mpi \alpha n^2}{3P} |q|.  \label{eq:q2q}
\end{equation}
(\ref{eq:q2q}) was found earlier in \cite{levitovsound}; the above formula is written in terms of simple thermodynamic quantities.  (\ref{eq:q2q}) recovers the well-known square root dispersion of plasmons in graphene \cite{hwangplasmon}, interpolating between propagating plasmons as $q\rightarrow 0$ and propagating sound modes as $q\rightarrow \infty$.   The long range nature of the Coulomb interactions, with $K\sim |q|^{-1}$, is responsible for the curious scaling of the second term of (\ref{eq:q2q}).

Let us note a few things about these results.  Firstly, by studying the dissipationless limit, it is easy to identify the diffusive mode (\ref{eq:coudiff}) as associated with the dynamics of the combination $\tilde P + 2\mpi \alpha n |q|^{-1} \tilde n$.   Secondly,  if $n=0$, then no matter how strong the Coulomb interactions are, the dispersion relation reduces to that of simple sound modes.   At finite density, the excitations look like plasmons whenever \begin{equation}
\omega \lesssim \frac{4\mpi \alpha n^2}{3P} \sim \frac{\alpha \mu^2}{T}\;\;\;\;(\mu \ll T).
\end{equation}  
Above this frequency, the propagating modes will look very similar to ordinary sound.   However, this crossover is quite slow.

\begin{figure}[t]
\centering
\includegraphics[width=7in]{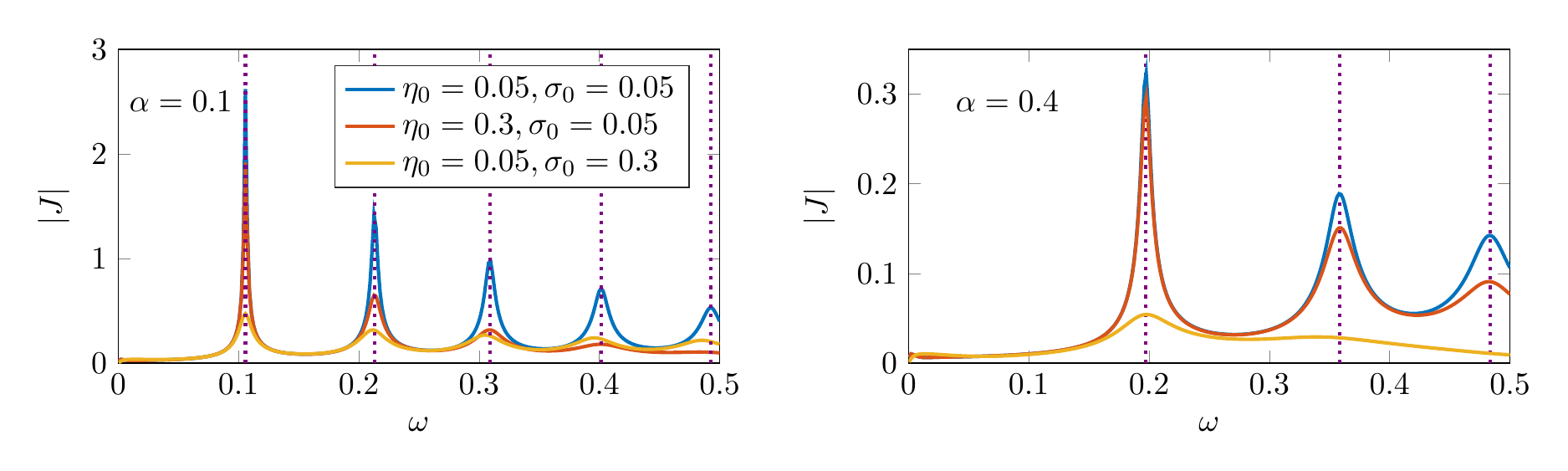}
\caption{A plot of $|J|$ vs. $\omega$ for various values of $\eta_0$ and $\sigma_0$, keeping $\Gamma=1$, $C_0=C_2=1$, $\mu=0.6$, $\gamma=0$, $x_0=L/2$ and $L=50$ fixed.  Vertical dotted lines denote the predicted resonances of (\ref{eq:q2q}) when $q=n\mpi/L$, for $n=1,3,5,\ldots$.}
\label{cleancoulombfig}
\end{figure}

We show in Appendix \ref{sec:appcouNM} that normal modes are \emph{exactly} present in our finite domain, with (\ref{eq:2paq}) obeyed exactly when $q=n\mpi/L$ with $n$ a positive integer.   Hence, we predict that in our set-up,  at any odd $n$ there is a normal mode in the dissipationless limit when $x_0=L/2$, so we expect to see sharp resonances in $J$ at the associated values of $\omega$, predicted by (\ref{eq:q2q}).   In Figure \ref{cleancoulombfig}, we numerically show this is the case.

\section{Disorder}\label{sec:dis}
In this section, we will study the role of disorder on the propagation of sound modes, and the observation of resonances.   Analogous to \cite{lucas, lucas3}, we assume that disorder is introduced via inhomogeneity in the chemical potential: \begin{equation}
\mu(x) = \bar\mu_0 + \sum_{n=1}^{N} \mu_n  \sin\left(\frac{n\mpi x}{L}+\phi_k\right),
\end{equation}
with $\mu_k$ and $\phi_k$ random variables.   $\bar\mu_0$ denotes the average value of the chemical potential, and $N$ counts the number of disorder modes that are included.  Such disorder is externally imposed on the fluid via charged impurities, and is known to be the dominant source of disorder in graphene, for example \cite{yacoby2007}.   We denote with $u$ the strength of disorder, defined through the variance of the chemical potential profile: \begin{equation}
u^2 \equiv \mathbb{E}\left[(\mu-\bar\mu_0)^2\right] \equiv \frac{1}{L}\int\limits_0^L \mathrm{d}x\; (\mu(x)-\bar\mu_0)^2.  \label{eq:udef}
\end{equation}
In graphene, the local $\mu(x)$ does not fluctuate with heavy-tailed statistics \cite{yacoby2007}.  For this reason, we approximate that $\mu(x)$ consists of a sum of sine waves with random coefficients, as in \cite{lucas3}.   We assume that the $\mu_k$ are independent and identically distributed with a uniform distribution, chosen so that (\ref{eq:udef}) holds.   The disorder correlation length $\xi$ is defined by \begin{equation}
\xi \equiv \frac{L}{N}.   \label{eq:xidef}
\end{equation}

For simplicity in this section, we usually assume that the fluid is on average charge neutral (to avoid, as much as possible, the transition to plasmons described in the previous section).   Hence, we will set $\bar\mu_0=0$ unless otherwise stated.

Disorder has a variety of interesting consequences.  One of them can be observed already at $\omega=0$.   Before, we showed that there is no signal in our set-up at $\omega=0$,   but in the presence of disorder, we generally find a finite (though small) value of $|\tilde J(x=0,\omega=0)|$ -- see Appendix \ref{app:DCanal}.

\subsection{Analytic Results}\label{sec:dis1}
Recall that $x_0$ is the distance between our source of energy and (one) boundary.   The measurable signal at the boundary is generically exponentially small in $x_0$:  \begin{equation}
|\tilde J(x=0)| \sim \exp\left[-\frac{x_0}{\xi_{\mathrm{loc}}}\right].  \label{eq:xilocdef}
\end{equation}
$\xi_{\mathrm{loc}}$ corresponds to the length scale over which the fluid effectively responds to our localized injection of energy,  and it is now our goal to compute of $\xi_{\mathrm{loc}}$ in a disordered fluid.   This allows us to understand what the limiting effects are for observing signatures of sound waves in experiments.   It is also of interest from a theoretical viewpoint:  we will discover multiple time scales between which the hydrodynamic response of the disordered electron fluid is dominated by entirely different processes.     

For now, we will neglect long range Coulomb interactions, which allows us to derive some analytic results for $\xi_{\mathrm{loc}}$.  We will later justify this assumption when $\bar\mu_0=0$ numerically.

So far, dissipative effects for the electronic fluid (viscosity, electron-phonon coupling) have been entirely responsible for  finite $\xi_{\mathrm{loc}}$.   In the clean limit, \begin{equation}
\frac{1}{\xi_{\mathrm{loc}}} = \left| \mathrm{Im}(q_{\mathrm{s}})\right|.
\end{equation}
For simplicity, let us focus on what happens as $\omega \rightarrow 0$.   If $\gamma =0$ then \begin{equation}
\frac{1}{\xi_{\mathrm{loc}}} = \frac{\sqrt{d}\eta\omega^2}{2(d+1)P}.  \label{eq:xiloc4}
\end{equation}
If $\gamma \ne 0$, we instead find 
\begin{equation}
\frac{1}{\xi_{\mathrm{loc}}} = \sqrt{\frac{d\omega\gamma}{2(d+1)P}}.  \label{eq:xiloc3}
\end{equation}
In this subsection, we address how $\xi_{\mathrm{loc}}$ is modified in the presence of disorder, both as $\omega \rightarrow 0$ and at higher frequencies.   The computations of $\xi_{\mathrm{loc}}$ are much more involved in disordered fluids.  In some limiting cases, we may compute $\xi_{\mathrm{loc}}$ analytically.   These computations can be found in Appendix \ref{app:loc} -- we summarize the results here and discuss the physical interpretations.   We  also caution the  experimentally-oriented reader that many of the parameters we employ in numerical simulations in this subsection are unrealistic for experiments -- the numerical parameters are chosen to make quantitative contact with simple theoretical predictions.   Some of these parameters can be made more realistic using the rescaling symmetries (\ref{eq:rescale}).

\begin{figure}[t]
\centering
\includegraphics[width=5.5in]{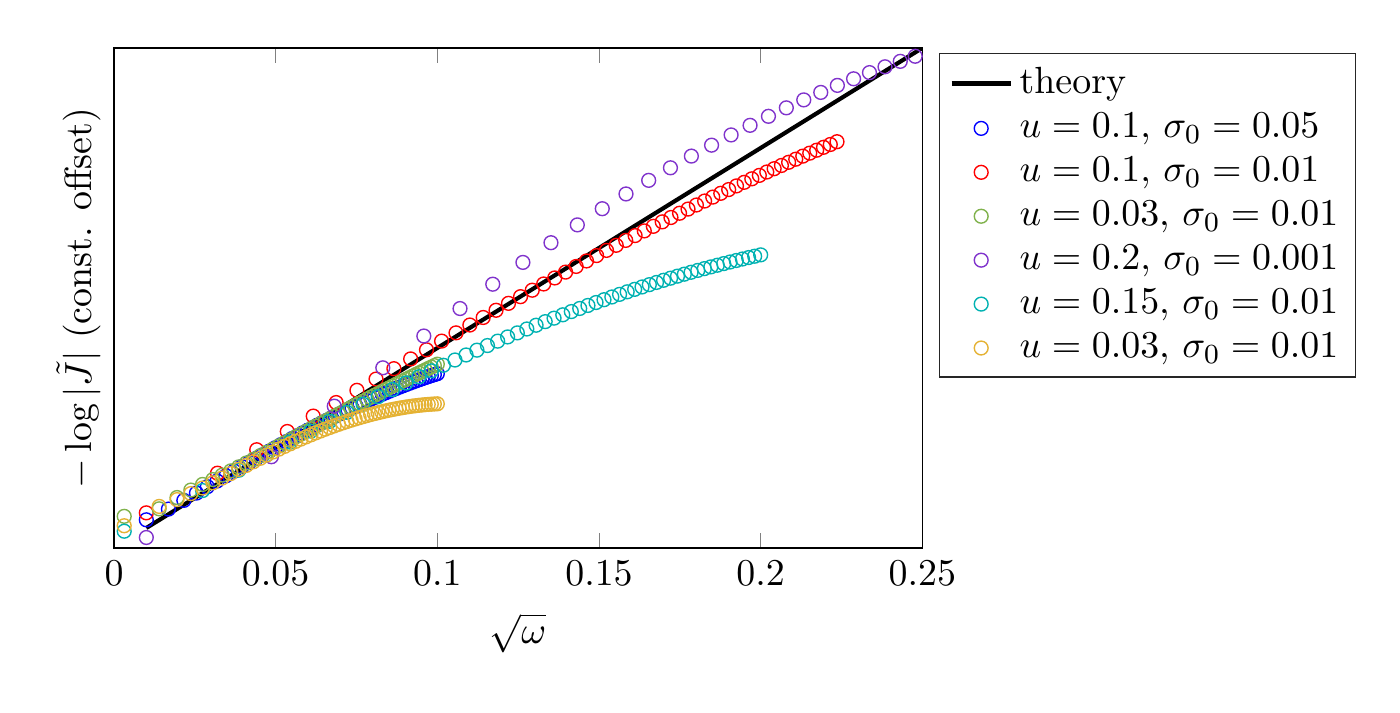}
\caption{Numerical simulations measuring $\xi_{\mathrm{loc}}$, as defined by (\ref{eq:xilocdef}).  We compare to the theoretical prediction given in (\ref{eq:xiloc2}), as $\omega \rightarrow 0$.  Each set of data points corresponds to a single disorder realization.   All simulations used $d=2$, $\eta_0=0.1$, $C_0=C_2=1$, $L\sim 1000$ (precise values vary, but were unimportant),  and $\alpha=\gamma = 0$.   The $y$-axis has been rescaled for each curve in accordance with theoretical predictions, and the only fit parameter in numerics is the overall amplitude of $\tilde J$.   Hence, there is a constant offset in the vertical direction, but the slope of the lines is not a fit parameter.  The scaling behavior ends when $\omega \ll \tau_{\mathrm{cp}}^{-1}$, the predicted momentum relaxation rate:  for the simulations displayed,  $\tau_{\mathrm{cp}}^{-1} \gtrsim T$.}
\label{fig:loc2}
\end{figure}
 
As $\omega \rightarrow 0$, assuming $\bar\mu_0=0$, we generically find \begin{equation}
\frac{1}{\xi_{\mathrm{loc}}} = \sqrt{\frac{d\omega}{2\sigma_{\textsc{q}}(\epsilon+P)}\left(\frac{\partial n}{\partial \mu}\right)^2 u^2}.  \label{eq:xiloc2}
\end{equation}
Figure \ref{fig:loc2} demonstrates that (\ref{eq:xiloc2}) is indeed found in numerical simulations.  We can understand (\ref{eq:xiloc2}) as follows.   The particular power of $\omega$ which appears in (\ref{eq:xiloc2}) is consistent with \emph{homogeneous} diffusive transport on the longest length scales, analogous to the case of momentum relaxation due to coupling with acoustic phonons (\ref{eq:omegadiff}).   Indeed, we may interpret (\ref{eq:xiloc2}) as \begin{equation}
\frac{1}{\xi_{\mathrm{loc}}} = \sqrt{\frac{d\omega}{\tau_{\mathrm{cp}}}},   \label{eq:tcp2}
\end{equation}
where $\tau_{\mathrm{cp}}$ is the momentum relaxation rate of the fluid due to the inhomogeneous chemical potential \cite{lucas3}.\footnote{In \cite{lucas3}, $\tau_{\mathrm{cp}}$ was computed by studying the thermoelectric conductivities, which characterize the response of the fluid to uniform electric fields and temperature gradients.   That these two different computations lead to the same $\tau_{\mathrm{cp}}$ justifies our interpretation.}    (\ref{eq:tcp2}) holds even for $\bar\mu_0 \ne 0$, and is derived in (\ref{eq:tcp}).   Upon comparing to (\ref{eq:xiloc3}) it is natural to postulate (as is done almost uniformly in the recent literature on hydrodynamics of electron fluids) that (\ref{eq:linear}) is a good model of dynamics in disordered media,  with $\gamma$ accounting for momentum relaxation due to all channels -- both electron-phonon coupling and disorder:  \begin{equation}
\gamma_{\mathrm{eff}} \equiv \frac{\epsilon+P}{\tau_{\mathrm{eff}}} \equiv \gamma_{\mathrm{el-ph}} + \frac{\epsilon+P}{\tau_{\mathrm{cp}}}.  \label{eq:gammasum}
\end{equation}  
Indeed, it was shown perturbatively in \cite{lucas, lucas3} that for the purposes of computing dc transport (in higher dimensions as well), we may approximate that the fluid is exactly homogeneous, but with a correction to $\gamma$ due to the disorder in the chemical potential.  This can be understood on rather general grounds at the quantum mechanical level \cite{lucasMM}.  

However, this ``relaxation time approximation" is not always accurate for spatiotemporal dynamics at finite frequency:  \begin{itemize}
\item Even at rather small values of $\omega$.   (\ref{eq:omegadiff}) and (\ref{eq:gammasum}) predict that only when $\omega \tau_{\mathrm{cp}} \gtrsim 1$ will we see the breakdown of (\ref{eq:xiloc2}).  In fact, Figure \ref{fig:loc2} shows the breakdown of the $\xi_{\mathrm{loc}} \sim \omega^{-1/2}$ scaling when $\omega \tau_{\mathrm{cp}} \sim 0.01$, in some cases.   Hence, at finite frequency the approximation (\ref{eq:gammasum}) will fail.  
\item We predicted in (\ref{eq:expsupgamma})  that (\ref{eq:gammasum}) implies that $|\tilde J|$ is always exponentially suppressed,  even when $\omega\tau_{\mathrm{cp}} \gg 1$.   In Figure \ref{fig:2dis},  we will show that the exponential  suppression of resonances is far less pronounced for actual disordered systems.
\end{itemize}

Why does the approximaton (\ref{eq:gammasum}) fail?   An intuitive argument  is as follows:  (\ref{eq:xiloc2}) appears independent of the disorder correlation length $\xi$, defined in (\ref{eq:xidef}).   But if we send a finite frequency sound mode through the fluid, we expect the response to be different depending on whether or not $\omega \xi \ll 1$ or $\omega \xi \gg 1$.   In the latter regime,  the wavelength of sound is very small compared to $\xi$, and so locally the medium wil look completely homogeneous.    The WKB approximation suggests that once $\omega \xi \gg 1$,  a sound mode should hardly feel the effects of the disordered background.

This intuition is almost -- but not exactly -- correct.   Neglecting dissipative effects in the equations of motion,  when $\omega \xi \sim 1$, waves may become classically Anderson localized.\footnote{In truth, waves are always localized in low dimensions, but when $\omega \xi \gg 1$ the localization length is essentially negligible for practical purposes.}    As a wave phenomenon, Anderson localization has been studied in a variety of disordered classical wave equations \cite{anderson2, john};  the localization of classical sound has even been observed experimentally in solids \cite{tiggelen}.    The breakdown of (\ref{eq:gammasum}) will be due to the onset of a regime where Anderson localization dominates the response.

\begin{figure}[t]
\centering
\includegraphics[width=7in]{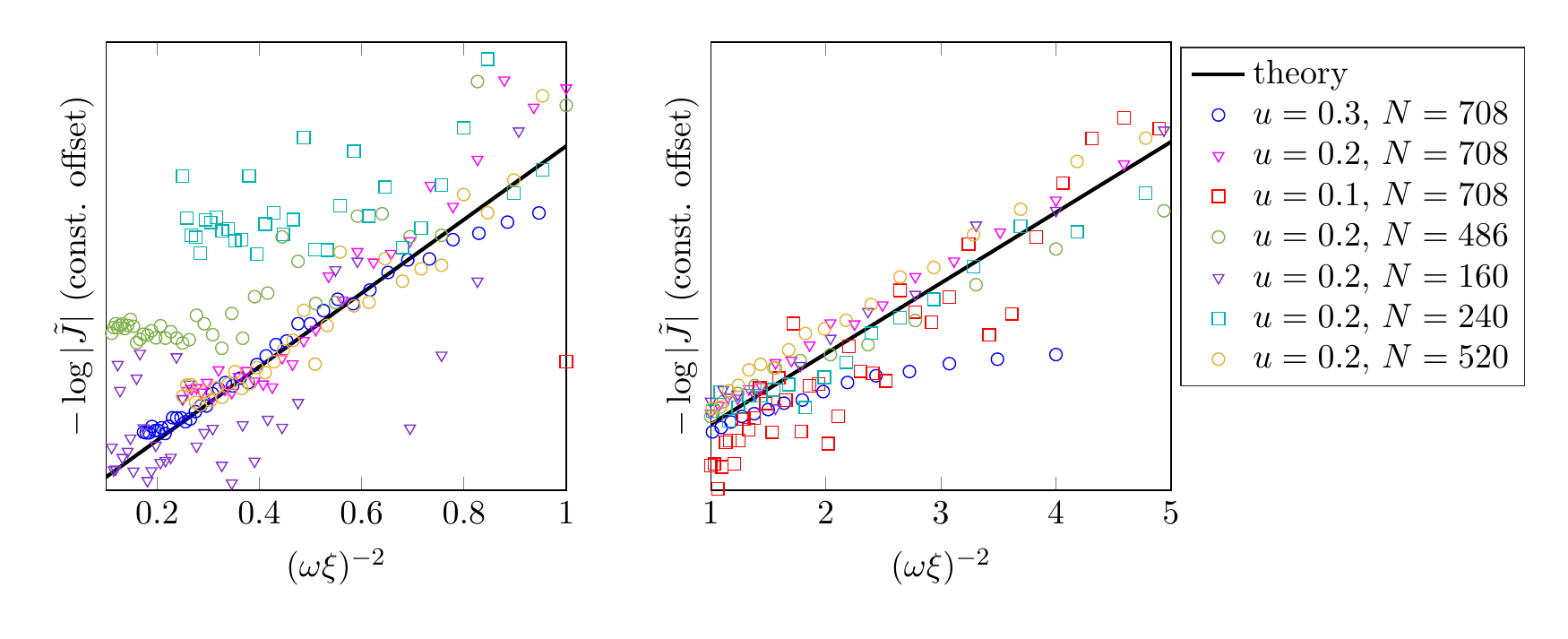}
\caption{Numerical simulations measuring $\xi_{\mathrm{loc}}$, as defined by (\ref{eq:xilocdef}).  We compare to the theoretical prediction given in (\ref{eq:xilocmain}), which is the localization length for sound waves.    We have split the data into two plots, to emphasize the scaling at slightly larger values of $\omega\xi$.    All simulations used $d=2$, $\sigma_0=\eta_0=10^{-3}$, $C_0=C_2=1$, $L\sim 2000$ (precise values vary, but were unimportant),  and $\alpha=\gamma = 0$.   The $y$-axis has been rescaled for each curve in accordance with theoretical predictions, and the only fit parameter in numerics is the overall amplitude of $\tilde J$.   Hence, there is a constant offset in the vertical direction, but the slope of the lines is not a fit parameter. Each set of data corresponds to a single realization of disorder;  the ``rugged" nature of some curves is related to high frequency resonances that are pronounced.  The agreement between numerics and theory is insensitive to the specific values of $\sigma_0$ and $\eta_0$, so long as they are small.  Changing $N$ corresponds to changing $\xi$, through (\ref{eq:xidef}).}
\label{fig:loc1}
\end{figure}

(\ref{eq:xiloc4}) suggests that when $\eta$ is small,  if $\gamma=0$ then there is a very large window over which the dissipative decay of sound waves should be negligible (at high frequency).  So let us focus on the (for now, heuristic) limit where $\eta,\sigma_{\textsc{q}} \rightarrow 0$ and $\xi \rightarrow \infty$.   In this limit,  $\omega \sim 1/\xi$ should be small enough so that viscous dissipation is relatively weak.  Neglecting dissipative ($\eta$, $\gamma$, $\sigma_{\textsc{q}}$) terms in the hydrodynamic equations, we find in Appendix \ref{app:loc} that the dynamics reduces to a simple inhomogeneous wave equation.  Once $\omega \xi \lesssim 1$, a calculation reveals that
\begin{equation}
\frac{1}{\xi_{\mathrm{loc}}} \approx \frac{2d\mpi^4}{5} \left(\frac{\partial n}{\partial \mu}\right)^2 \frac{ u^4}{(Ts\omega)^2 \xi^3}.  \label{eq:xilocmain}
\end{equation}
For the first time, we see that  $\xi_{\mathrm{loc}}$ can be finite even in a dissipationless\footnote{Here, we are using the term dissipationless to refer to the fact that entropy production vanishes.   Energy and momentum can, however, be exchanged via coupling to the external chemical potential.} fluid, so long as there is disorder.  The interpretation of this is natural.  (\ref{eq:xilocmain}) is the Anderson localization length of classically localized sound waves.   In contrast to (\ref{eq:xiloc2}),  (\ref{eq:xilocmain}) predicts that the decay of the signal now decreases as $\omega$ increases.   Figure \ref{fig:loc1} demonstrates (\ref{eq:xilocmain}) can be observed in numerical simulations.     

We have seen that the localization length of sound modes can control the response of our fluid to a localized perturbation at frequencies $\omega\xi \sim1 $, and that this response is effectively dissipationless.   It is rather strange, then, that the response given by (\ref{eq:xiloc2}) as $\omega \rightarrow 0$ is a dissipative response (as $\xi_{\mathrm{loc}}$ depends on $\sigma_{\textsc{q}}$).   To understand how this can occur, we first note that the dissipationless hydrodynamic equations become \emph{ill posed} at $\omega=0$:  we find two conservation laws (entropy and charge) which are inconsistent with one another when $\sigma_{\textsc{q}} =0$ \cite{lucas3}.   Hence, dissipation is necessary to even have a well-posed set of equations.   More physically, the presence of conservation laws means that the full, dissipative hydrodynamic equations of motion must have delocalized modes at $\omega=0$, corresponding to the transport of conserved quantities.    These delocalized modes suggest that the true localization length must diverge as $\omega \rightarrow 0$, in a manner which is consistent with (\ref{eq:xiloc2}) \cite{ziman2, amir2}.   Indeed, when (\ref{eq:xiloc2}) holds,  the localization length of the eigenstates of the hydrodynamic equations may generically grow faster than $\omega^{-1/2}$, and will not generically equal $\xi_{\mathrm{loc}}$.   This is acceptable: at low frequencies, dissipative processes -- and not wave interference -- are responsible for the spatial decay of the signal which is measured by $\xi_{\mathrm{loc}}$.

We thus find that disordered electron fluids are rather interesting.    At both low and high frequency, hydrodynamic dissipation ($\sigma_{\textsc{q}}$ and $\eta$) plays a crucial in the spatiotemporal dynamics.   However, there may be a broad range of intermediate frequencies,  near $\omega \xi \sim 1$,  where dissipative effects become rather negligible.   In sharp contrast to clean fluids, where ideal hydrodynamics emerges on very long time scales,  ideal hydrodynamics is best probed in an electron fluid at strictly finite frequencies.

\begin{figure}[t]
\centering
\includegraphics[width=4in]{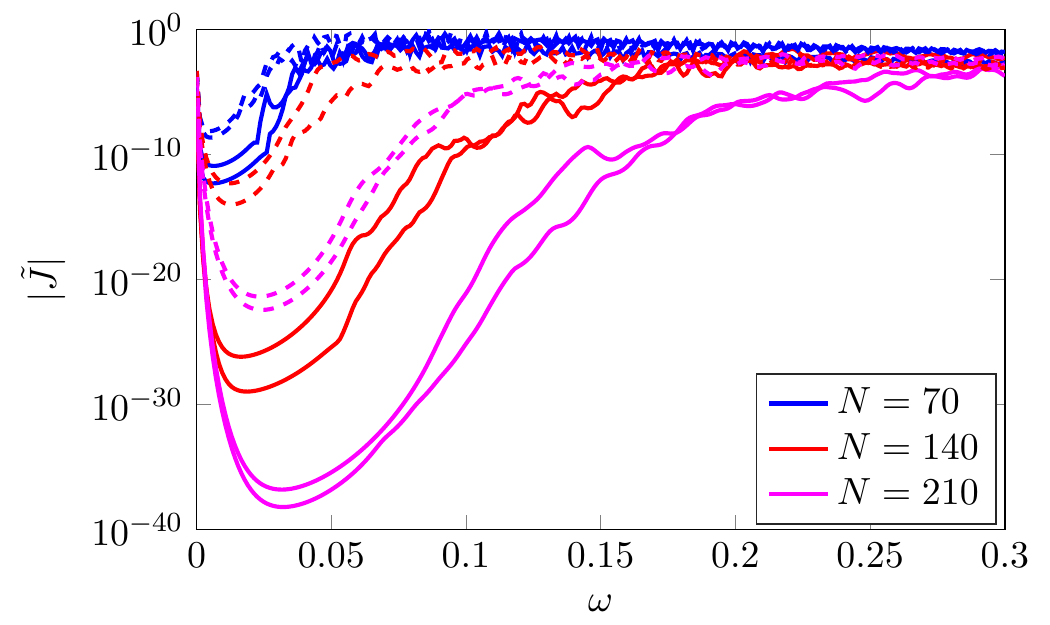}
\caption{A plot of $|\tilde J|$ vs. $\omega$ for various disorder realizations.   We used parameters $L=1000$, $\sigma_0=\eta_0=0.01$, $C_0=C_2=1$, $\gamma=0$.  Dashed lines have $u=0.1$ and solid lines have $u=0.2$.    Note in particular how all solid lines overlap at small frequencies, independent of $N$ (and thus $\xi$), in agreement with (\ref{eq:xiloc2}). The transition between (\ref{eq:xiloc2}) and (\ref{eq:xilocmain}) is qualitatively visible.  We also caution that data points with $|\tilde J| \lesssim 10^{-30}$ may be unreliable as our numerical computation employs double-precision arithmetic.}
\label{fig:omegaxi}
\end{figure}

So far, the discussion of the difference between (\ref{eq:xiloc2}) and (\ref{eq:xilocmain}) has been qualitative.   In Figure \ref{fig:omegaxi},  we show numerically the transition from (\ref{eq:xiloc2}) to (\ref{eq:xilocmain}) and beyond.    In the appendix we discuss this transition a bit more quantitatively.   The punchline is as follows:  the momentum relaxation time approximation (\ref{eq:xiloc2}) is quantitatively accurate for \begin{equation}
\omega \lesssim  \omega_{\mathrm{bd}} \equiv  \mathcal{A}_0 \frac{\sigma_{\textsc{q}}}{\xi^2},   \label{omegafail}
\end{equation}
where $\mathcal{A}_0$ is a dimensionful quantity, dependent on the thermodynamic equations of state, but seemingly not on $u$.   We expect that above this scale,  but below $\omega \xi \sim 1$,  (\ref{eq:xilocmain}) properly describes the response.    Since (\ref{omegafail}) scales as $\xi^{-2}$,  as $\xi \rightarrow \infty$,  there can be a parametrically large regime of frequencies where the classical localization of sound waves dominates the non-local response in our set-up.    Interestingly, the localized response is larger than the response predicted by (\ref{eq:xiloc2}).

The theory that we have discussed is particular to one-dimensional disorder.   As is well-known,  localization becomes much weaker in higher dimensions \cite{abrahams}.   This one-dimensional theory is probably not relevant to experiments, except in very long and thin samples.     Nonetheless, the qualitative signatures of Anderson localization may show up in experiments, as we will detail in the next subsection.

\subsection{Numerical Results}
\begin{figure}[t]
\centering
\includegraphics[width=6.6in]{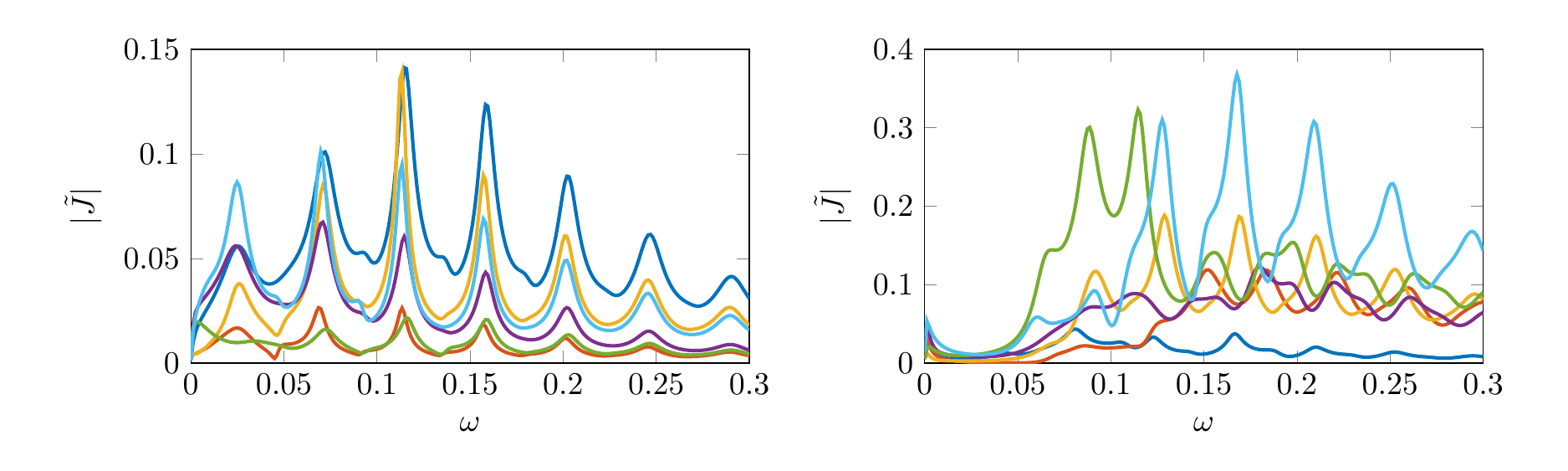}
\caption{A plot of $|\tilde J|$ vs. $\omega$ for 6 disorder realizations.   We used parameters $\bar\mu_0=0$, $N=8$, $L=100$, $\sigma_0=\eta_0=0.1$, $C_0=1$, $\gamma=0$, $x_0=L/2$.   Left panel: $C_2=0.3$;  right panel:  $C_2=1$.   In the left panel, we see a generic alignment of higher order resonances in $\omega$ between disorder realizations.   In the right panel, we see hints of localizaton at small frequencies, and spurious resonances at higher frequencies.}
\label{fig:2dis}
\end{figure}

Figure \ref{fig:omegaxi} compares the signals between two different disorder realizations;  on a logarithmic scale, we see that for the large samples being studied,  the difference between disorder realizations is qualitatively rather minor.    However, keeping in mind the relevant length scales for graphene, we note that $L=1000$ is about a factor of 10 larger than what is currently feasible.   Hence, we now turn to numerical simulations in smaller samples, where finite size effects are important, to discuss the practical feasibility of our set-up.    Figure \ref{fig:2dis} compares the response 6 disorder realizations for samples with a smaller amount of disorder.   We see that depending on the ratio $C_2/C_0$,   the response is more or less sensitive to the particular disorder realization, especially as $\omega$ increases.    At smaller values of $C_2/C_0$ we see that higher order resonances ($n=5,7,\ldots$) occur at frequencies similar to the clean theory,  whereas at higher values some samples either do not have visible resonances, or these resonances do not appear at the same frequencies.

\begin{figure}[t]
\centering
\includegraphics[width=7in]{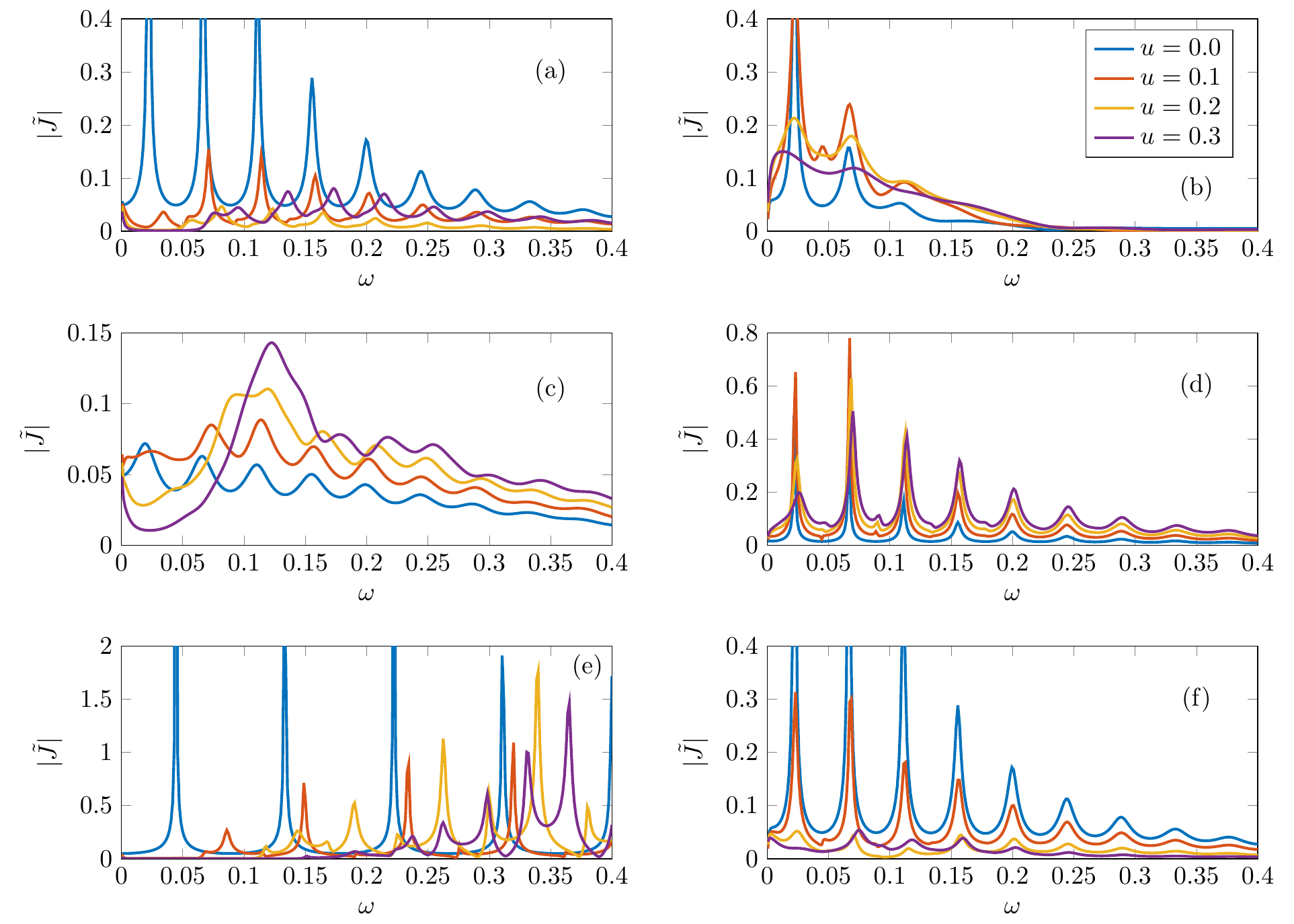}
\caption{A plot of $|\tilde J|$ vs. $\omega$ for $\bar\mu_0=0.1$ and a single disorder realization, but with the amplitude of disorder rescaled.   Parameters are generically as follows:  $\alpha=0$, $C_0=C_2=1$, $\gamma=0$, $x_0=L/2$, $\eta_0=\sigma_0=0.1$, with the following exceptions:  (b) $\eta_0=\sigma_0=1$;  (c) $\gamma=0.02$;  (d) $C_2=0.3$;  (e) $\sigma_0=\eta_0=0.01$;  (f) $\sigma_0=1$.    Extensive comments on this figure may be found in the main text. }
\label{fig:incdis}
\end{figure}

In Figure \ref{fig:incdis},  we demonstrate the effects of increasing disorder strength for a large variety of different possible equations of state.   In each case, we assume that $\bar\mu_0=0$.  Below we comment on general lessons: \begin{itemize}
\item We see that when $\sigma_0$ is small enough, the effects of localization become quite severe.   This follows from the fact that $1/\xi_{\mathrm{loc}} \sim \sigma_0^{-1/2}$ in (\ref{eq:xiloc2}).   This is especially visible in (e),  which is reminiscent of Figure \ref{fig:omegaxi}.    As $\sigma_0$ increases, we first observe the effects of localization becoming less and less important, albeit visible in (a), and the main challenge is that resonances are shifted away from their ideal frequencies.   Once $\sigma_0$ is large enough, then higher order resonances are not quantitatively shifted from their predicted frequencies.
\item The dominant effect of viscosity near charge neutrality determines the sharpness of resonances, as well as the number of visible resonances: compare panels (a, b, f).
\item As we emphasized in this section, momentum relaxation due to electron-phonon coupling is qualitatively distinct at higher frequencies, compared to momentum relaxation due to disorder.   This is most evident in panel (c),  where we see that the dominant effect of electron-phonon scattering, encoded in $\gamma$, is simply to reduce the overall signal $\tilde J$, as well as further broaden the resonances.   Other qualitative features of (c) are shared with panel (a), which has identical parameters up to $\gamma=0$:   shifting of resonances at large disorder, and hints of localization at stronger disorder.
\item As in Figure \ref{fig:2dis},  we see that when $C_2 \sim \partial n/\partial \mu$ is smaller, the signal becomes much cleaner, at all disorder strengths:  see panel (d).
\end{itemize}

\begin{figure}[t]
\centering
\includegraphics[width=7in]{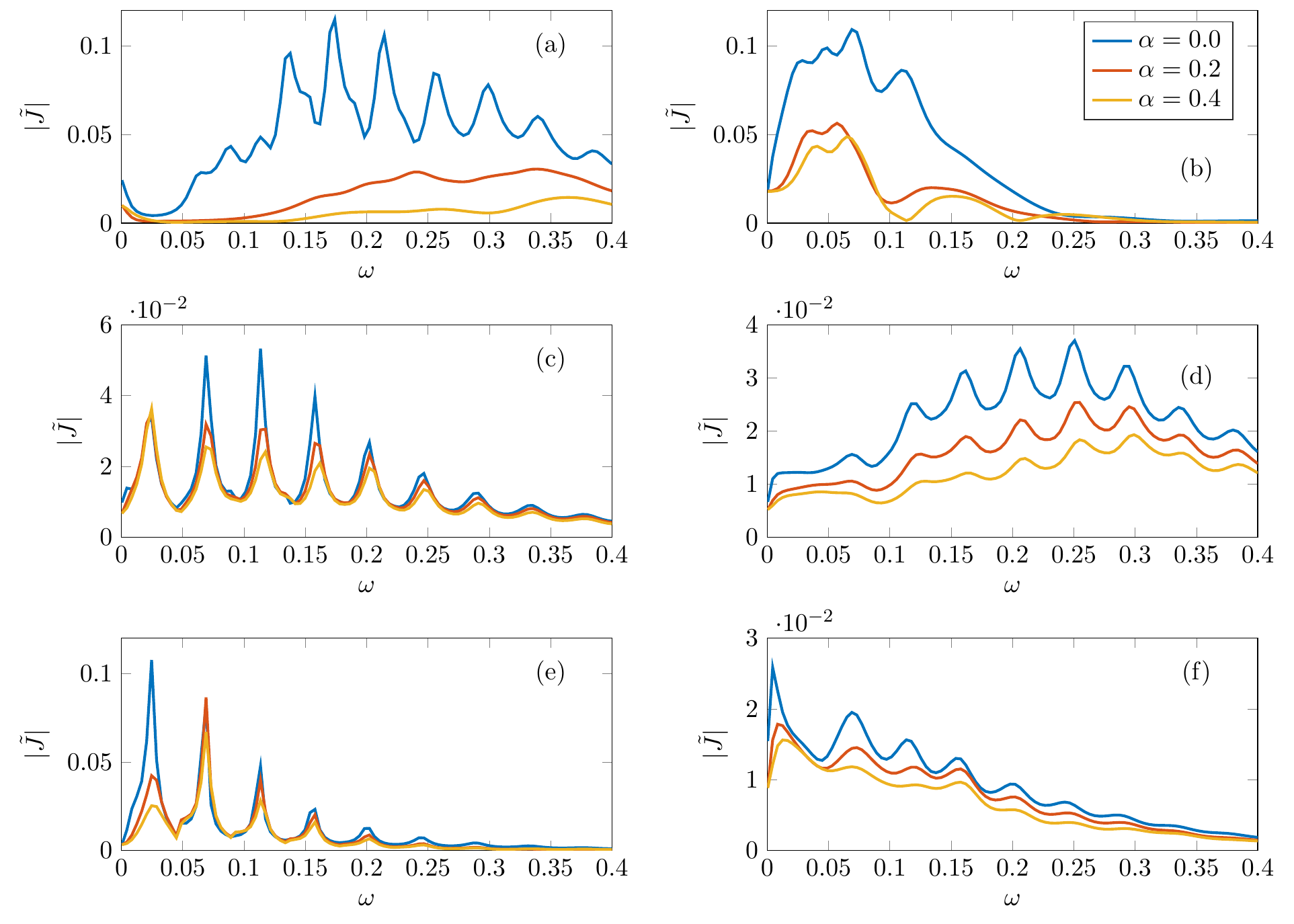}
\caption{A plot of $|\tilde J|$ vs. $\omega$ for $\bar\mu_0=0$ and a single disorder realization, but with the amplitude of $\alpha$ rescaled.   Parameters are generically as follows: $N=8$,  $C_0=1$, $C_2=0.3$, $\gamma=0$, $x_0=L/2$, $\eta_0=\sigma_0=0.1$, $u=0.2$, with the following exceptions:  (a,b) $C_2=1$;  (b) $\eta_0=\sigma_0=1$;  (d) $N=20$;  (e) $u=0.1$;  (f) $\gamma=0.02$.}
\label{fig:incalpha}
\end{figure}

Figure \ref{fig:incdis} neglects the effects of long range Coulomb interactions.   In Figure \ref{fig:incalpha},  we demonstrate the consequences of long-range Coulomb interactions in disordered samples.   The key points are as follows: \begin{itemize}
\item Depending on $C_2$,  the effects of Coulomb interactions when disorder is taken into account can either nearly destroy resonances (a,b),  or have relatively minor effects (c-f).   In the Dirac fluid, it is predicted (see (\ref{eq:dfpredict})) that $C_2/C_0$ is closer to 0.3,  suggesting that the effects of Coulomb screening may not be as severe in experiment.
\item If finite $\alpha$ not outright destroy resonances, as in panels (c-f),  then they do not shift substantially the locations of a given resonance.   Indeed if disoder is weak, as in panel (e),  it is possible for the signal to be practically immune to $\alpha$.
\item In panels (c-f), increasing $\alpha$ decreases the amplitude of the observable signal,  though this effect rarely destroys resonances outright.
\item Perhaps surprisingly, some of the features of localization (lack of low $\omega$ resonances) are visible irrespective of the strength of  non-local Coulomb interactions.   This effect is most evident in panels (a,d).
\end{itemize}

\section{Outlook}
In this paper, we have theoretically demonstrated that charge neutral strongly interacting electron-hole plasmas that arise in ultrapure solid state systems, such as the Dirac fluid in graphene, have sharply defined sound modes that can be observed upon injecting energy into the system at a modulated rate.  At the charge neutrality point,  we have emphasized that the acoustic resonances are essentially immune to any possible long-range Coulomb screening,  and are also robust to moderate amounts of disorder.

\subsection{Experimental Viability of AC Response in Graphene}
Is the detection of such acoustic resonances experimentally plausible in graphene?   As we noted earlier, our setup is somewhat similar to the set-up in \cite{song2, ma1, song, johannsen, ma2}, where a laser locally injects energy into the sample.   The crucial difference in this paper is that energy is not injected at a constant rate,  but instead at a finite (angular) frequency $\omega$.

The most important question to ask is whether or not it is feasible to drive the electronic system in the linear response regime.    Here we make a crude estimate of this.    The total energy rate injected into the system is given by \begin{equation}
\Gamma_{\mathrm{tot}} = \Gamma W \sim \hbar \omega_{\mathrm{EM}} \times \alpha_{\mathrm{EM}}
\end{equation}
where $W$ is the transverse width of the sample, $\omega_{\mathrm{EM}}$ is the (angular) frequency of laser light, and $\alpha_{\mathrm{EM}}$ is the rate at which photons are absorbed.   In principle, we require $\alpha_{\mathrm{EM}} \gtrsim \omega$,  where $\omega$ is the modulation frequency of the signal.    Assuming $W\sim 10$ $\mmu$m,  $\omega \sim 100$ GHz and $\omega_{\mathrm{EM}} \sim$ 1000 THz, we obtain \begin{equation}
\frac{\Gamma}{\epsilon+P} \sim 10^4 \; \frac{\mathrm{m}}{\mathrm{s}} \sim 10^{-2}\times v_{\mathrm{F}}.
\end{equation}
Hence the velocity of the excited fluid is much smaller than $v_{\mathrm{F}}$ and indeed the velocity of the electronic fluid should be small.     At an extremely sharp resonance, linear hydrodynamics would break down -- however, this is acceptable since there would nevertheless be strong evidence for a resonance.

The most serious challenge is to detect the rapid changes in the electric current at angular frequency $\omega$.   Plasmons have been detected at $\omega \gg 1$ THz  \cite{chen12, koppens12, koppens1601},  but such high frequencies are beyond the regime of validity of hydrodynamics.     It may also be easier to study the second order response to a finite frequency drive,  which will contain an $\omega=0$ component.   This static response will be easier to detect experimentally, given the challenges with driving the electronic fluid at $\omega \lesssim 1$ THz.

It may also be possible to use optical transmission experiments to detect signatures of hydrodynamics (see \cite{zaanen} for a theoretical model in a Fermi liquid).   However, here the ``bumps" in the signal are exponentially suppressed, in contrast to the resonances in Figure \ref{fighydroclean}.

\subsection{Open Theoretical Questions}
The main theoretical development presented in this paper is the analysis of non-local finite frequency response of disordered electron-hole plasma.    With or without Coulomb interactions, we see that the finite frequency response can behave qualitatively differently from the response predicted by ``momentum relaxation time" theories,  at parametrically smaller frequencies than predicted by $\tau_{\mathrm{eff}}^{-1}$, as given  in (\ref{eq:gammasum}).   Hence, these toy models of disorder can exhibit qualitatively different finite frequency response than other models of disorder.   This almost certainly has relevance for many other quantum systems,  including those studied using holographic duality, where similar cartoons of disorder which neglect inhomogeneity are quite popular \cite{vegh, donos1, andrade}.   We expect the dynamical response of disordered electron fluids to hence be much more interesting -- especially on intermediate time scales -- than previous studies would suggest.   It would be interesting to investigate this further in future work.

Deep in the Fermi liquid regime, it is possible to crisply use non-local electrical measurements to detect viscous electron flow \cite{polini, levitovhydro, bandurin}.    However, near charge neutrality the energy-momentum and charge sectors decouple within hydrodynamics (up to charged disorder).  Indeed, \cite{bandurin} found no experimental evidence for viscous electron flow at charge neutrality.   In this work, we have focused on finite frequency response as a way to detect hydrodynamics near charge neutrality, but it is also important to develop tests for time-independent hydrodynamic flows near charge neutrality.    Thermoelectric transport has been one proposal \cite{hkms, crossno, lucas3}, but there may be a non-local measurement which provides complementary evidence for hydrodynamics. 

Although we have focused on graphene in relating our theory to  experiment,  there are other materials of interest which are charge neutral,  including recently realized Weyl semimetals \cite{soljacic, syxu, bqlv}.   It would be of interest to observe sound resonances in any such electron-hole plasma.  

Finally, it is of great interest to move beyond the linear response regime when studying hydrodynamic electron flow in solid-state systems.   Indeed, experimental observation of nonlinear hydrodynamic phenomena such as turbulence would be remarkable.  However, electron fluids are complicated by the breaking of boost invariance by Coulomb interactions, mediated by the true speed of light $c$,   and so the equations of state may pick up a complicated velocity dependence.\footnote{This phenomenon seems to be hinted at in \cite{sodemann}, and may be an entirely quantum mechanical effect.}   Whether or not these effects have a qualitative change on nonlinear flow patterns is an important question for future study.

\addcontentsline{toc}{section}{Acknowledgements}
\section*{Acknowledgements}
I would like to thank Richard Davison, Kin Chung Fong, Qiong Ma and Subir Sachdev for helpful discussions.   This work was supported by the NSF under Grant DMR-1360789 and MURI grant W911NF-14-1-0003 from ARO.
\begin{appendix}

\section{Dimensional Analysis}\label{app:dimanal}
Here we present the relevant energy, length, etc. scales for graphene, assuming temperature $T\sim 100$ K:  \begin{subequations}\begin{align}
\text{velocity} &\sim v_{\mathrm{F}} \sim 10^6 \; \mathrm{m/s}, \\
\text{length} &\sim \frac{\hbar v_{\mathrm{F}}}{k_{\mathrm{B}}T} \sim 100\; \mathrm{nm}, \\
\text{time} &\sim \frac{\hbar }{k_{\mathrm{B}}T} \sim 0.1\; \mathrm{ps}, \\
\text{energy} &\sim k_{\mathrm{B}}T \sim 100 \; \mathrm{K} \sim 0.01 \; \mathrm{eV}, \\
\text{conductivity} &\sim \frac{e^2}{\hbar} \sim 0.25 \; \mathrm{k\Omega}^{-1}.
\end{align}\end{subequations}
For example, we frequently plot the response in the main text as a function of frequency.  $\omega=0.1$ refers to $\omega = 0.1 \times 10^{13} \; \mathrm{s}^{-1} = 10^{12} \; \mathrm{s}^{-1}$.   We also note that chemical potential scales as energy, and charge density scales as $[\text{length}]^{-d}$, in our notation.

For the Dirac fluid, kinetic theory predicts that at leading order in $\alpha$ \cite{vafek, schmalian, muller4, muller3}: \begin{subequations}\label{eq:dfpredict}\begin{align}
C_0 &\approx 3.44 \alpha^2, \\
C_2 &\approx 0.88 \alpha^2, \\
\sigma_0 &\approx 0.12 \alpha^{-2},\\
\eta_0 &\approx 0.45.
\end{align}\end{subequations}

\section{Boundary Conditions}\label{app:bc}
Here we justify the claim in the main text that the hydrodynamic equations are well-posed with Dirichlet boundary conditions only on $\tilde \mu$ and $\tilde T$.    More precisely, we will show that 4 boundary conditions are sufficient to fix the problem.    To do so, let us consider a junction between two domains $R_-=[x_-,x_0)$ and  $R_+=(x_0,x_+]$ where we have a solution to the differential equations in the domains $R_-$ and $R_+$ separately.     We will show that there are two linearly independent constraints relating $\tilde \mu$, $\tilde T$, $\tilde v$ and their first derivatives.

Let us begin by supposing that $\partial_x \mu (x_0) = 0$.    Using that all parameters in the hydrodynamic equations are functions of $\mu$,  we conclude that all parameters  in (\ref{eq:linear}) are locally constant at $x_0$.  Then we obtain a constraint equation \begin{equation}
(\epsilon+P) \partial_x v = \mathrm{i}\omega d\left(n\tilde \mu + s\tilde T\right) + \mathcal{S}_0 .
\end{equation}
Upon plugging this equation into (\ref{eq:linearb}) we obtain \begin{equation}
n\partial_x \tilde \mu + s\partial_x \tilde T  - \frac{\eta^\prime}{\epsilon+P} \partial_x \left[  \mathrm{i}\omega d\left(n\tilde \mu + s\tilde T\right) + \mathcal{S}_0 \right] = \left[\mathrm{i}\omega (\epsilon+P) - \gamma \right]\tilde v,
\end{equation}
which provides us a linearly independent constraint equation.    At this point, we can see that we may locally remove both $\tilde v$ and $\partial_x \tilde v$, and are left with two independent second order equations for $\tilde \mu$ and $\tilde T$;  hence we have exhausted our constraints.

If $\partial_x \mu \ne 0$ locally, then we follow a similar procedure, beginning with the first constraint: \begin{equation}
\left[n\tilde v - \sigma_{\textsc{q}}\left(\partial_x \tilde \mu - \frac{\mu}{T} \partial_x \tilde T\right)\right] = \frac{\partial_x \left((\epsilon+P)\tilde v\right) - \mathcal{S}_0 - \mathrm{i}\omega d\left( n \tilde \mu + s \tilde T \right)}{\partial_x \mu}.
\end{equation} 
Now, taking the derivative of both sides, we obtain \begin{equation}
\mathrm{i}\omega \left(\frac{\partial n}{\partial \mu} \tilde \mu + \frac{\partial n}{\partial T} \tilde T\right)  = \partial_x \left[\frac{\tilde v\partial_x (\epsilon+P) + - \mathcal{S}_0 - \mathrm{i}\omega d\left( n \tilde \mu + s \tilde T \right)}{\partial_x \mu}\right] + \frac{\epsilon+P}{\partial_x \mu} \partial_x^2 \tilde v
\end{equation}
(\ref{eq:linearb}) gives us a linearly independent expression for $\partial_x^2 \tilde v$ in terms of $\tilde \mu$, $\tilde T$, $\tilde v$ and their first derivatives.   Hence, we find a second linearly independent constraint, just as before.   Again, this set of constraints is exhaustive, and the remaining equations are independent second order equations for $\tilde \mu$ and $\tilde T$.

A heuristic argument for the reason that there are 4 independent modes to these equations from the fact that there are a pair of sound modes and an independent diffusive mode.   Each such mode would correspond to one second order differential equation, implying that in total there are 4 boundary conditions to fix.

\section{Numerical Methods}\label{app:num}
We solve differential equations numerically using standard pseudospectral methods \cite{trefethen}.  We discretize the spatial coordinate $x$ into a list of $N$ points $\lbrace x_j\rbrace = \mathbf{x}$.  We further discretize \begin{equation}
\tilde \mu(\mathbf{x}) \approx \sum_{\alpha=0}^{N-1} \tilde \mu_\alpha  \mathrm{T}_\alpha\left(\frac{x}{L}\right)  = \sum_{i=1}^N \tilde \mu_i  f_i(x),
\end{equation} where $\tilde \mu_\alpha$ are coefficients,  $\mathrm{T}_\alpha$ is the Chebyshev polynomial of order $\alpha$, and $\tilde \mu_i \equiv \tilde \mu(x_i)$.   Exactly at the grid points of interest,  the discretized vector is chosen to agree with the exact function; away from these points, we interpolate the true function with a sum of polynomials.   Defining the $3N\times 1$ vector \begin{equation}
\mathbf{u} \equiv (\tilde \mu(x_1),\ldots,\tilde\mu(x_N),\tilde T(x_1),\ldots\tilde T(x_N),\tilde v(x_1),\ldots,\tilde v(x_N))^{\mathsf{T}}
\end{equation}
and definining an appropriate $N\times N$ (nonlocal) derivative matrix $\mathsf{D}$, acting on $\tilde \mu_i$ to approximate $\partial_x \tilde \mu(x_i)$,  differential equations of the form (\ref{eq:linear}) may then be written as \begin{equation}
\mathsf{L} \mathbf{u} = \mathbf{s}.   \label{eq:Lus}
\end{equation}
with $\mathsf{L}$ a $3N\times 3N$ matrix which is built as follows:  local functions such as $n(\mu)$ are turned into diagonal matrices with entries $n(\mu(x_j))$,  and $\partial_x$ is replaced by $\mathsf{D}$.

It is straightforward to implement $K\otimes \tilde n$ numerically.   $K$ becomes  a nonlocal matrix acting on a discrete vector of data $\tilde n_j = \tilde n(x_j)$.   Hence, as a discretized matrix, the Coulomb kernel $K$ becomes \begin{equation}
\mathsf{K}_{ij} \approx \frac{x_{j+1}+x_{j-1}-2x_j}{2} K(x_i,x_j),
\end{equation}
with straightforward modifications for the endpoints.

\subsection{Domain Decomposition}

For more complicated problems without long-range Coulomb interactions, we employ domain decomposition \cite{domaindec} to greatly increase the number of grid points in the computational domain.    The essential idea is as follows:   we divide our total grid \begin{equation}
[0,L] = \left[0,\frac{L}{N_{\mathrm{d}}}\right] \cup \left[\frac{L}{N_{\mathrm{d}}},2\frac{L}{N_{\mathrm{d}}}\right] \cup \cdots \cup \left[L-\frac{L}{N_{\mathrm{d}}},L\right].
\end{equation}
In each domain, we solve an equation of the form (\ref{eq:Lus}), subject to the boundary conditions $\tilde \mu$ and $\tilde T$ fixed but \emph{arbitrary} at the endpoints; hence, we solve the equation 4 times in each domain, to account for all possible linearly independent sets of boundary conditions.

We then must glue the domains together.  We demand continuity of $\tilde v$ and $\partial_x\tilde\mu$, which we have found to be more numerically stable at extremely low frequencies than continuity of $\partial_x \tilde T$ and $\partial_x\tilde\mu$.   This leads to a second linear algebra problem which we may solve for $\tilde \mu$ and $\tilde T$ at all (interior) endpoints, as $\tilde \mu(0,L)= \tilde T(0,L)=0$.   Once we have fixed the interior values of $\tilde \mu$ and $\tilde T$,  we may glue the solutions in each domain together to find a global solution.

\section{Galilean Invariant Fluids}\label{sec:appgal}
In this appendix we briefly review the theory of sound waves in ordinary Galilean invariant fluids \cite{forster}.   Galilean invariance imposes a slightly different form of the constitutive relations than (\ref{eq:hydrost}).   In particular, it is no longer appropriate to think about a relativistic stress-energy tensor -- such a tensor would no longer be isotropic, as the momentum density and energy current are distinct.    The linearized conservation laws for ``charge", energy and momentum read: \begin{subequations}\label{eq:galhydro}\begin{align}
\partial_t \tilde n + n \partial_x \tilde v &= 0, \\
mn\partial_t \tilde v + \partial_x \tilde P - \eta^\prime \partial_x^2 \tilde v &=0, \\
\partial_t \tilde\epsilon + (\epsilon+P) \partial_x \tilde v - \kappa_{\textsc{q}} \partial_x^2 \tilde T &= 0.
\end{align}\end{subequations}
In many cases -- such as the classical fluid water -- the ``charge" conservation listed above is not conservation of electric charge.   However, Galilean invariance affords an additional conserved quantity, which we commonly take to be particle number.   In the above equations $m$ is a mass parameter, and $\kappa_{\textsc{q}}$ is a dissipative coefficient.

We emphasize that $\kappa_{\textsc{q}}$ is distinct from $\sigma_{\textsc{q}}$.   In a relativistic fluid, the energy current is proportional to the momentum current,  but in a Galilean-invariant fluid the particle current is proportional to the momentum current.   This has important consequences for dissipation, as we will shortly see.

It is straightforward to construct the dispersion relations associated with (\ref{eq:galhydro}).   One finds a diffusive mode where $\tilde P \approx 0$ and \begin{equation}
\omega \approx - \mathrm{i} D k^2
\end{equation}
and a pair of sound modes: \begin{equation}
\omega \approx \pm c k  - \mathrm{i}\Gamma_{\mathrm{s}}k^2,
\end{equation}with \begin{subequations}\begin{align}
c^2 &= \frac{1}{m} \left(\frac{\partial P}{\partial n}\right)_{s/n}, \\
D &= \frac{\kappa_{\textsc{q}}}{Tn (\partial_T(s/n))_P} , \\
\Gamma_{\mathrm{s}} &= \frac{\eta^\prime}{2mn} + \frac{\kappa_{\textsc{q}}}{2} \frac{(\partial_n T)_{s/n}(\partial_\epsilon P)_n}{(\partial_n P)_{s/n}}  = \frac{\eta^\prime}{2mn} + \frac{\kappa_{\textsc{q}}}{2Tn}\left[\left(\frac{\partial T}{\partial (s/n)}\right)_n - \left(\frac{\partial T}{\partial (s/n)}\right)_P \right].
\end{align}\end{subequations}
As standard, thermodynamic derivatives with objects outside parentheses imply that the derivative is taken with the external quantity held fixed.  The thermodynamic identity \begin{equation}
Tn\mathrm{d}\frac{s}{n} = \mathrm{d}\epsilon - \frac{\epsilon+P}{n} \mathrm{d}n.
\end{equation}
is helpful to simplify the resulting expressions above.

There are two points worth emphasizing, to contrast with the theory of sound waves in the gapless relativistic fluids:  \begin{itemize}
\item A gapped relativistic fluid (or more generally, a relativistic fluid with a third energy scale beyond $\mu$ and $T$) will also have $\Gamma_{\mathrm{s}}$ not proportional to $\eta^\prime$.  $\Gamma_{\mathrm{s}}\sim \eta^\prime$ for ``relativistic" electron-hole plasma, where measuring sound attenuation can directly measure the viscosity.
\item In \cite{levitovsound},  the dissipative structure of the above equations was used (the energy current had a dissipative contribution).   While this hydrodynamics is suitable for Galilean invariant fluids, we see that the charge neutral ($n\rightarrow 0$) limit of (\ref{eq:galhydro}) is very singular, with the momentum and charge currents identically vanishing.   It is necessary to use the hydrodynamics presented in this paper for a consistent theory of sound in charge neutral fluids such as the Dirac fluid in graphene.
\end{itemize}

\section{Driving a Clean Fluid}\label{app:anal}

Recall from (\ref{eq:sound}) that the dynamics of $\tilde P$ and $\tilde v$ -- the only variables sourced by $\mathcal{S}$ -- decouple from $\tilde n$ in a clean fluid.  Keeping $\omega$ fixed, the solution to the wave equations is\begin{equation}
\tilde P =  \left\lbrace\begin{array}{ll} \displaystyle A \sin (\omega x/v_\eta)   &\  0<x<x_0 \\ \displaystyle B \sin (\omega (L-x)/v_\eta)    &\ x_0<x<L  \end{array}\right.,
\end{equation}
as we impose boundary conditions at $x=0$ and $x=L$ are that the electronic fluid does not exchange momentum with the leads.   Indeed, one can readily see from (\ref{eq:s4b}) in position space that $\tilde P = 0$ implies $\partial_x \tilde v = 0$, and hence the perturbed stress tensor will vanish at the boundaries.   Further integrating over the equations at $x=x_0$,  we find that \begin{equation}
\tilde v(x_0^+) - \tilde v(x_0^-) = \frac{\Gamma}{(d+1)P},  \label{eq:vjump}
\end{equation}
and that $P$ is continuous at $x_0$.   Using that \begin{equation}
\partial_x P = \tilde v \left[\mathrm{i}\omega(d+1)P - \gamma - \eta q^2_{\mathrm{s}}\right],  \label{eq:dxPv}
\end{equation}
we fix \begin{subequations}\begin{align}
A &= - \frac{\Gamma}{(d+1)Pq_{\mathrm{s}}\sin(q_{\mathrm{s}} L)}\left[\mathrm{i}\omega(d+1)P- \gamma - \eta q_{\mathrm{s}}^2\right] \sin(q_{\mathrm{s}} (L-x_0)), \\
B &= -  \frac{\Gamma}{(d+1)Pq_{\mathrm{s}}\sin(q_{\mathrm{s}} L)}\left[\mathrm{i}\omega(d+1)P -\gamma  -  \eta q_{\mathrm{s}}^2\right] \sin (q_{\mathrm{s}}x_0).
\end{align}\end{subequations}
Hence, the velocity field is given by \begin{equation}
\tilde v(x) = \left\lbrace \begin{array}{ll} \displaystyle -\frac{\Gamma}{(d+1)P}\dfrac{\sin(q_{\mathrm{s}}(L-x_0))}{\sin(q_{\mathrm{s}} L)} \cos(q_{\mathrm{s}}x)  &\ 0<x<x_0 \\ \dfrac{\Gamma}{(d+1)P}\dfrac{\sin(q_{\mathrm{s}} x_0)}{\sin(q_{\mathrm{s}} L)} \cos(q_{\mathrm{s}}(L-x))  &\ 0<x<x_0  \end{array}\right..
\end{equation}

As we mentioned in the main text,  $\tilde n$ is slave to $\tilde P$ in a sound wave.  The precise relation is:\begin{equation}
\left(\mathrm{i}\omega - Dq_{\mathrm{s}}^2\right) \tilde n = n \partial_x \tilde v + C \partial_x^2 \tilde P = \left(n + \frac{\mathrm{i}Cq_{\mathrm{s}}^2 (d+1)P}{d\omega}\right)   \partial_x \tilde v.
\end{equation}
Hence, in a sound wave the electric current is given by \begin{equation}
\tilde J =  \left(n + \frac{\mathrm{i}Cq_{\mathrm{s}}^2 (d+1)P}{d\omega}\right) \frac{\mathrm{i}\omega}{\mathrm{i}\omega - Dq_{\mathrm{s}}^2} \tilde v .  \label{eq:Jinv}
\end{equation}
(\ref{eq:vjump}) implies that $\tilde J$ is not continuous at $x=x_0$, which violates the equations of motion.   Hence, we must add to our solution a charge diffusion mode which makes the net $\tilde J$ continuous at $x=x_0$.   This diffusive mode (which only involves $\tilde n$) takes the form \begin{equation}
\tilde n(x) = c\times \left\lbrace \begin{array}{ll}  \sin(q_{\mathrm{d}}(L-x_0)) \sin(q_{\mathrm{d}} x)  &\  0<x<x_0 \\ \sin(q_{\mathrm{d}}x_0)\sin(q_{\mathrm{d}}(L-x)) &\ x_0<x<L  \end{array}\right.
\end{equation}
where $c$ is a constant to be determined.   Note that, as required, $\tilde n$ is continuous at $x=x_0$.   Employing (\ref{eq:Jinv}),  that the diffusive mode above creates electric current $\tilde J = -D\partial_x\tilde n$, and requiring continuity of the net $\tilde J$ at $x=x_0$ fixes \begin{equation}
q_{\mathrm{d}}D \sin (q_{\mathrm{d}}L) c = \left(n + \frac{\mathrm{i}Cq_{\mathrm{s}}^2 (d+1)P}{d\omega}\right) \frac{\mathrm{i}\omega}{\mathrm{i}\omega - Dq_{\mathrm{s}}^2} \frac{\Gamma}{(d+1)P}.
\end{equation}
From here, it is straightforward to obtain (\ref{eq:jmain}).

\section{Coulomb Kernel in a Finite Domain}\label{sec:appcou}
\begin{figure}[t]
\centering
\includegraphics[width=3.7in]{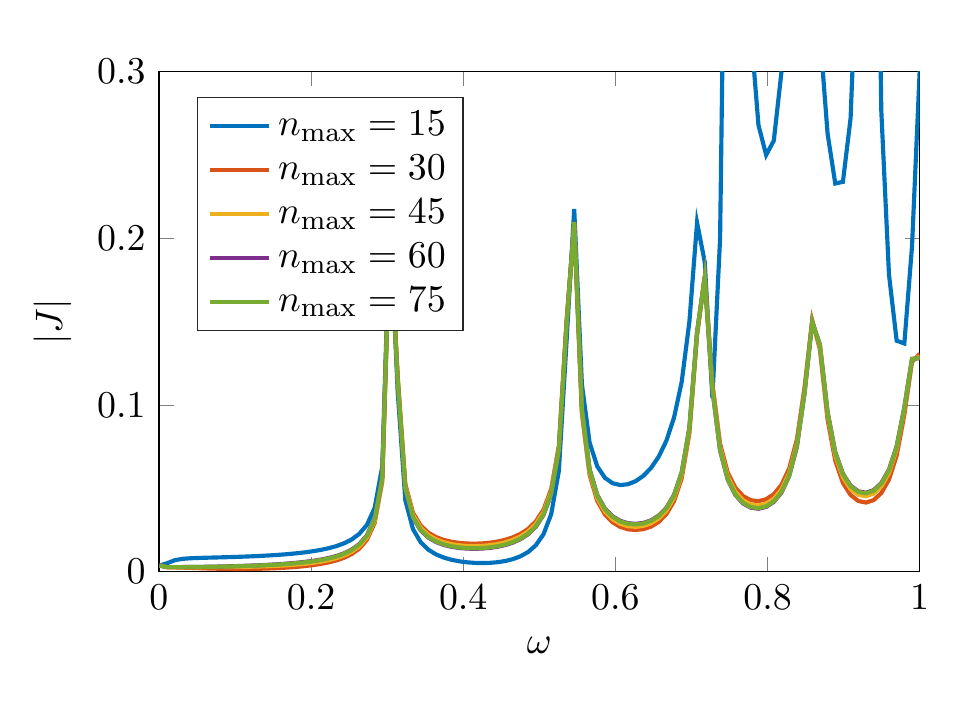}
\caption{A comparison of $|J|$ as a function of $\omega$, for more precise estimates of the Coulomb kernel $K(x,y)$ (parameterized by $n_{\mathrm{max}}$).   We show the response in a clean plasma with the parameters $L=50$, $\alpha=1$, $C_0=C_2=1$, $\eta_0=\sigma_0=0.01$, and $\bar\mu_0 = 0.6$, with 571 grid points.   Rapid convergence is observed so long as $n_{\mathrm{max}} \gtrsim 50$. }
\label{figconvg}
\end{figure}

In a finite domain, the Coulomb kernel $K(x,y)$, defined via \begin{equation}
\tilde\varphi(x) =  \int \mathrm{d}y K(x,y)\tilde n(y)
\end{equation}
is not translation invariant.   We construct it by solving (\ref{eq:poisson}) in a Fourier series,  assuming a point source $\tilde n = \mdelta(x-y)$: \begin{equation}
\left(\partial_x^2 + \partial_z^2 \right) \tilde\varphi = -4\mpi \alpha \mdelta(x-y)\mdelta(z).  \label{eq:2delta}
\end{equation}
$K(x,y)$ is then given by $\tilde\varphi(x,z=0)$.   Away from $z=0$ we may solve (\ref{eq:2delta}) in a Fourier series: \begin{equation}
\varphi = \sum_{n=1}^\infty \varphi_n \mathrm{e}^{-n\mpi |z|/L} \sin \frac{n\mpi x}{L}.
\end{equation}
Integrating (\ref{eq:2delta}) over $z=0$, we find that \begin{equation}
-\sum_{n=1}^\infty \frac{2n\mpi}{L}\varphi_n \sin \frac{n\mpi x}{L}= -4\mpi \alpha \mdelta(x-y),
\end{equation}
from which we find \begin{equation}
\frac{2n\mpi}{L} \varphi_n = \frac{2}{L}\int\limits_0^\infty \mathrm{d}x\;  4\mpi \alpha  \mdelta(x-y) \sin \frac{n\mpi x}{L} = \frac{8\mpi \alpha}{L} \sin\frac{n\mpi y}{L},
\end{equation}
so we conclude that \begin{equation}
K(x,y) = \lim_{n_{\mathrm{max}} \rightarrow\infty} \sum_{n=1}^{n_{\mathrm{max}}} \frac{4\alpha}{n}  \sin \frac{n\mpi x}{L} \sin\frac{n\mpi y}{L}.
\end{equation}
In our numerics, we evaluate $K(x,y)$ by keeping $n_{\mathrm{max}} \gg 1$ finite.   In Figure \ref{figconvg}, we show that our calculation of the acoustic response  is not sensitive to $n_{\mathrm{max}}$ once $n_{\mathrm{max}} \gtrsim 50$.   All plots in the main text use $n_{\mathrm{max}} = 200$.

\subsection{Normal Modes}\label{sec:appcouNM}
Here we justify the claim that in the main text,  this finite-domain Coulomb kernel nonetheless supports the same plasmonic modes of the infinite plane, subject to the constraint $q=m\mpi /L$, for $m\in\mathbb{N}$.  To do this, we simply note that \begin{equation}
\int\limits_0^L \mathrm{d}y \; K(x,y) \sin\frac{m\mpi y}{L} = \int\limits_0^L \mathrm{d}y \; \sum_{n=1}^\infty \frac{4\alpha}{n}  \sin \frac{n\mpi x}{L} \sin\frac{n\mpi y}{L}  \sin\frac{m\mpi y}{L} = \frac{2\mpi \alpha}{m\mpi L^{-1}} \sin \frac{m\mpi x}{L},
\end{equation}
as all terms in the sum over $n$ vanish by orthonormality, except for the term where $m=n$.   Hence, we see that $K\otimes \tilde n \sim \tilde n$ as promised in the main text, with the proportionality coefficient given in (\ref{eq:2paq}).

\section{Analytic DC Result}\label{app:DCanal}
In this appendix we analytically compute $\tilde J$ in the case $\omega=0$.    Assuming $\mathcal{S}(x)=0$ locally, both the charge $\tilde J$ and heat current $\tilde Q$ are exactly conserved \cite{lucas3}. giving us a simple expression for the velocity \begin{equation}
\tilde v = \frac{\tilde Q}{\epsilon+P} + \frac{\mu \tilde J}{\epsilon+P}.
\end{equation}
We have the remaining two equations \begin{subequations}\begin{align}
\frac{Ts}{\epsilon+P}\tilde J - \frac{n}{\epsilon+P}\tilde Q &= -\sigma_{\textsc{q}} \partial_x \tilde \mu + \sigma_{\textsc{q}} \frac{\mu}{T} \partial_x \tilde T, \\
\partial_x \left(\eta\partial_x \frac{\mu}{\epsilon+P}\right) \tilde J + \partial_x \left(\eta\partial_x \frac{\tilde Q }{\epsilon+P}\right) &=  n \partial_x \tilde \mu + s\partial_x \tilde T. 
\end{align}\end{subequations}
Upon solving these equations we obtain \begin{subequations}\begin{align}
\partial_x \tilde \mu &= \tilde J \left[\frac{\mu}{\epsilon+P} \partial_x \left(\eta\partial_x \frac{\mu}{\epsilon+P}\right) - \frac{(Ts)^2}{\sigma_{\textsc{q}}(\epsilon+P)^2}\right] + \frac{Tsn\tilde Q}{\sigma_{\textsc{q}}(\epsilon+P)^2} + \frac{\mu}{\epsilon+P} \partial_x \left(\eta\partial_x \frac{\tilde Q}{\epsilon+P}\right), \\
\partial_x \tilde T &= \left[\frac{T^2sn}{\sigma_{\textsc{q}}(\epsilon+P)^2} + \frac{T}{\epsilon+P}\partial_x \left(\eta\partial_x \frac{\mu}{\epsilon+P}\right)\right]\tilde J + \left[ \frac{T}{\epsilon+P} \partial_x \left(\eta\partial_x \frac{\tilde Q}{\epsilon+P}\right)- \frac{Tn^2\tilde Q}{\sigma_{\textsc{q}}(\epsilon+P)^2}\right].
\end{align}\end{subequations}
Assuming $\mathcal{S} = \Gamma \mdelta(x-x_0)$, then $\tilde J$ is a constant and \begin{equation}
\tilde Q(x) = \tilde Q(0)+ \Gamma \mathrm{\Theta}(x-x_0).
\end{equation}
Defining the spatially averaged heat current as \begin{equation}
\tilde Q_0 \equiv \Gamma \left(1-\frac{x_0}{L}\right) + \tilde Q(0),
\end{equation}
in the thermodynamic limit we find, upon demanding that $\tilde \mu(0) = \tilde \mu(L) = \tilde T(0) = \tilde T(L) = 0$: \begin{subequations}\begin{align}
\frac{\Gamma}{L}\left(\frac{\eta}{\epsilon+P} \partial_x \frac{\mu}{\epsilon+P}\right)_{x=x_0} &= -\tilde J \mathbb{E}\left[ \eta\left(\partial_x \frac{\mu}{\epsilon+P}\right)^2 + \frac{(Ts)^2}{\sigma_{\textsc{q}}(\epsilon+P)^2}\right] + \tilde Q_0 \mathbb{E}\left[\frac{Tsn}{\sigma_{\textsc{q}}(\epsilon+P)^2} - \eta \partial_x \frac{\mu}{\epsilon+P} \partial_x \frac{1}{\epsilon+P}\right] \\
\frac{\Gamma}{L}\left(\frac{\eta}{\epsilon+P} \partial_x \frac{1}{\epsilon+P}\right)_{x=x_0} &=  \tilde J \mathbb{E}\left[\frac{Tsn}{\sigma_{\textsc{q}}(\epsilon+P)^2} - \eta \partial_x \frac{\mu}{\epsilon+P} \partial_x \frac{1}{\epsilon+P}\right] -\tilde Q_0 \mathbb{E}\left[ \eta\left(\partial_x \frac{1}{\epsilon+P}\right)^2 + \frac{n^2}{\sigma_{\textsc{q}}(\epsilon+P)^2}\right] 
\end{align}\end{subequations}
where we denote $\mathbb{E}[\cdots]  = L^{-1}\int_0^L \mathrm{d}x\cdots$.   This set of two equations is straightforwardly solved for $\tilde J$.   In the weak disorder limit, we find  \begin{equation}
\tilde J \approx \frac{\Gamma \tau_{\mathrm{cp}}}{L}\left(\frac{d\eta n^2 \partial_x \mu}{((d+1)P)^3}\right)_{x=x_0}
\end{equation}
where we have defined $\tau_{\mathrm{cp}}$ in (\ref{eq:tcp}).

\section{Localization}\label{app:loc}
\subsection{Short Wavelengths}
We begin by studying the localization problem at short wavelengths (a notion which we will clarify in the course of the calculation).   For simplicity, we neglect the dissipative terms in the equations of motion, which themselves lead to the decay of sound modes -- our purpose here is to isolate the effects coming from classical Anderson localization (and hence beyond the ``mean field" descriptions of disorder which frequently appear in the literature).   The equations of motion (\ref{eq:linear}) are, without microscopic dissipation, \begin{subequations}\label{eq:appc}\begin{align}
0 &=  -\mathrm{i}\omega \tilde n + \partial_x (n\tilde v),  \label{eq:appc1} \\
n\tilde v \partial_x \mu &= -\mathrm{i}\omega d \tilde P + \partial_x ((d+1)P\tilde v), \\
\tilde n \partial_x \mu &=  -\mathrm{i}\omega (d+1)P\tilde v + \partial_x \tilde P.
\end{align}\end{subequations}
Defining the energy current \begin{equation}
\tilde \Pi \equiv (d+1)P \tilde v
\end{equation}
and the parameter \begin{equation}
G \equiv \frac{n}{(d+1)P},
\end{equation}
(\ref{eq:appc}) can be reduced to the equation \begin{equation}
\partial_x^2 \tilde \Pi + k^2 \tilde \Pi = (G \partial_x^2 \mu ) \tilde \Pi + ((d+1)\partial_x \mu )\partial_x(G\tilde \Pi)  \label{eq:appc2}
\end{equation}
where we have defined $k=\omega\sqrt{d}$, as usual.

Next, we employ a trick from \cite{lifshitz, lugan} in order to calculate the localization length of short wavelength (large $k$) sound waves described by (\ref{eq:appc2}).   To do so, we define the two functions $r(x)$ and $\theta(x)$ as \begin{subequations}\label{eq:rtheta}\begin{align}
\tilde\Pi &\equiv  r\sin\theta, \\
\partial_x \tilde\Pi &\equiv  kr\cos\theta.
\end{align}\end{subequations}
If there was no disorder,  then the solution to these equations would be $r=\text{constant}$ and $\theta = kx$.    Our goal will be to perturbatively construct a solution accounting for the disorder.  Upon plugging in (\ref{eq:rtheta}) into (\ref{eq:appc2}) we obtain \begin{subequations}\begin{align}
\frac{\partial_x r}{r} &=  \mathcal{U} \frac{1+\cos(2\theta)}{2}  + \frac{\sin(2\theta)}{2} \mathcal{V}, \\
\partial_x \theta - k &= -\frac{\mathcal{U}}{2} \sin(2\theta) - \frac{1-\cos(2\theta)}{2} \mathcal{V},
\end{align}\end{subequations}
where we have defined \begin{subequations}\begin{align}
\mathcal{U} &\equiv (d+1)G\partial_x \mu, \\
\mathcal{V} &\equiv \frac{(d+1)\partial_x \mu \partial_x G + G\partial_x^2\mu}{k}.  \label{eq:defmcv}
\end{align}\end{subequations}
At first order in $\mathcal{U}$ and $\mathcal{V}$, we find that \begin{equation}
\theta \approx kx - \frac{1}{2}\int\limits_0^x \mathrm{d}x^\prime \; \left[\sin(2kx^\prime)\mathcal{U}(x^\prime) + (1-\cos(2kx^\prime)) \mathcal{V}(x^\prime)\right] \equiv kx + \tilde \theta(x),
\end{equation}
and hence \begin{align}
\log \frac{r(x)}{r(0)} &\approx \int\limits_0^x \mathrm{d}x^\prime \left\lbrace \mathcal{U}(x^\prime) \left(\frac{1+\cos(2kx^\prime)}{2}-\sin(2kx^\prime) \tilde\theta(x^\prime)\right) + \left(\frac{\sin(2kx^\prime)}{2} +\cos(2kx^\prime)\tilde\theta(x^\prime) \right)\mathcal{V}(x^\prime)   \right\rbrace
\end{align}
We wish to isolate the terms in the above expression where the integrand is, on average as a function of $x$,  a constant.   In this case, we will find \begin{equation}
\log \frac{r(x)}{r(0)} \approx \frac{x}{\xi_{\mathrm{loc}}},
\end{equation} with $\xi_{\mathrm{loc}}$ the localization length of the Anderson localized waves.  The most obvious such term is $\mathcal{U}(x^\prime)$.   However, upon recalling the definition of $\mathcal{U}$,  we note that \begin{equation}
\mathcal{U}(x) = \frac{n}{P}\partial_x \mu = \frac{\partial \log P}{\partial \mu} \partial_x \mu = \frac{\partial \log P}{\partial x},
\end{equation}
and so in fact this term is a total derivative which will not contribute parametrically to the integral.   The dominant contributions to this integral come from expanding out $\tilde \theta$, and re-arranging terms.  For example, \begin{align}
&\mathcal{U}(x^\prime)\sin(2kx^\prime)\tilde\theta(x^\prime) = \notag \\
& -\frac{1}{2} \int\limits_{-x^\prime}^0 \mathrm{d}x^{\prime\prime}  \mathcal{U}(x^\prime)\sin(2kx^\prime) \left[ \mathcal{U}(x^\prime + x^{\prime\prime}) \sin(2k(x^\prime+x^{\prime\prime})) + (1-\cos(2k(x^\prime+x^{\prime\prime})))\mathcal{V}(x^\prime+x^{\prime\prime}) \right] \notag \\
&= -\frac{1}{4}\int\limits_{-x^\prime}^0\mathrm{d}x^{\prime\prime}\left[ \cos(2kx^{\prime\prime})\mathcal{U}(x^\prime)\mathcal{U}(x^\prime + x^{\prime\prime}) - \mathcal{U}(x^\prime)\mathcal{V}(x^\prime + x^{\prime\prime}) \sin(2kx^{\prime\prime}) \right]  + \mathrm{O}(\sin(x^\prime), \cos(x^\prime))
\end{align}
Collecting the pieces in which the integrand is independent of $x^\prime$, and assuming the disorder distribution is stationary (in $x$), we obtain \begin{equation}
\frac{1}{\xi_{\mathrm{loc}}} = \frac{1}{4}\int\limits_{-\infty}^0 \mathrm{d}u \left[\mathbb{E}\left[\mathcal{U}(0)\mathcal{U}(u) + \mathcal{V}(0)\mathcal{V}(u)\right]\cos(2ku) - \mathbb{E}\left[\mathcal{U}(0)\mathcal{V}(u) + \mathcal{V}(0)\mathcal{U}(u)\right]\sin(2ku)  \right]  \label{xiloc}
\end{equation}
The localization length of sound modes of frequency $\omega$ is inversely proportional to certain disorder autocorrelations at frequency $2\sqrt{d}\omega$.

To make further progress in analyzing (\ref{xiloc}), let us now specialize to the case where the fluid is overall charge neutral, and assume that $\omega \xi \ll 1$ and that $\mu(x)\ll T$.    In this limit, $\mathcal{U}$ may be neglected relative to $\mathcal{V}$.   We further assume that $\mu(x)$ takes the form \begin{equation}
\mu(x) \approx \frac{Ts_0}{\chi_0} G(x) = \sum_{n=-N}^N a_n \mathrm{e}^{2\mpi \mathrm{i} nx/L}, \label{eq:Fdis1}
\end{equation}
with $s_0$ and $\chi_0$ the entropy density and charge susceptibility $\chi = \partial_\mu n$ of the clean fluid, $a_0=0$ and $a_n$ zero mean (complex) Gaussian random variables obeying $\overline{a_n} = a_{-n}$ and \begin{equation}
\mathbb{E}_{\mathrm{d}}\left[a_n a_m\right] = u^2 \frac{\mdelta_{n,-m}}{2N}, \;\;\; (n\ne 0).  \label{eq:Fdis2}
\end{equation}
$\mathbb{E}_{\mathrm{d}}[\cdots]$ denotes averages over the quenched random variables $a_n$.  This normalization is chosen so that \begin{equation}
\mathbb{E}_{\mathrm{d}}\left[\mathbb{E}\left[ \mu(x)^2\right] \right] = \sum_{m,n=-N}^N \mathbb{E}_{\mathrm{d}}[a_m a_n]  \mathbb{E}\left[\mathrm{e}^{2\mpi \mathrm{i}(m+n)x/L}\right] = \sum_{n=-N}^N \frac{u^2}{2N} = u^2,
\end{equation}
consistent with our definition of $u$ in the main text.    Henceforth, we will usually be sloppy about distinguishing between $\mathbb{E}[\cdots]$ and $\mathbb{E}_{\mathrm{d}}[\cdots]$, which we expect are equivalent in the thermodynamic limit. Defining \begin{equation}
k_m \equiv \frac{2\mpi m}{L},
\end{equation}we obtain from (\ref{eq:defmcv}) \begin{equation}
\mathcal{V}(x) = -\frac{\chi_0}{Ts_0 \sqrt{d}\omega} \sum_{k_n,k_m}  \left[(d+1) k_n k_m + k_m^2 \right] a_n a_m \mathrm{e}^{\mathrm{i}(k_n+k_m)x}.
\end{equation}Using Wick's Theorem, and neglecting an overall constant contribution:\begin{align}
\mathbb{E}\left[\mathcal{V}(0)\mathcal{V}(x)\right] = \frac{\chi_0^2}{d(Ts_0\omega)^2} \sum_{k_n k_m} \frac{u^4}{8N^2} \left(2(d+1)k_mk_n + k_m^2 + k_n^2 \right)^2  \mathrm{e}^{\mathrm{i}(k_n+k_m)x} + \text{constant}
\end{align}
Next, consider for $k_l \ne 0$:\begin{align}
\frac{1}{\xi_{\mathrm{loc}}(k_l)} &\equiv \frac{1}{8}\int\limits_0^L \mathrm{d}x  \mathbb{E}\left[\mathcal{V}(0)\mathcal{V}(x)\right] \mathrm{e}^{\mathrm{i}k_l x}  \notag \\
&= \frac{L}{8N} \frac{\chi_0^2}{d(Ts_0\omega)^2} \frac{u^4}{8N} \sum_{k_m}\left(k_m^2 + (k_m+k_l)^2 -2(d+1)k_m(k_m+k_l)\right)^2.  \label{eq:58}
\end{align}
The spatial integral enforces $k_n = -k_m - k_l$.   Defining \begin{equation}
\xi \equiv \frac{L}{N},
\end{equation}
we note that when $\omega \xi \ll 1$,  $\omega \sim k_l$ and $k_l \ll k_N = 2\mpi / \xi$ and hence at leading order, (\ref{eq:58}) is independent of $k_l$ and approximately given by \begin{equation}
\frac{1}{\xi_{\mathrm{loc}}(0)} \approx \frac{\xi}{64N}\frac{\chi_0^2 u^4}{d(Ts_0\omega)^2} \sum_{k_m} (2dk_m^2)^2 \approx \frac{(2\mpi)^4}{40} \frac{d\chi_0^2 u^4}{(Ts_0\omega)^2 \xi^3}. 
\end{equation}
The result is independent of $L$, as it should be.  Hence, we obtain the expression for $\xi_{\mathrm{loc}}$ quoted in (\ref{eq:xilocmain}).

For $\omega \xi \gg 1$,  then we obtain a finite value of $\xi_{\mathrm{loc}}$ by either extending our perturbative calculation to higher orders,  or using very high order corrections to the equations of state, which enter the equation through $\mathcal{U}$ and $\mathcal{V}$ in non-trivial ways.   In particular,  $1/(\epsilon+P)$, and hence $\mathcal{U}$ and $\mathcal{V}$, will generally contain Fourier modes of arbitrarily large wave number, as it is not a polynomial of fixed order in $\mu(x)$.   We expect such effects to be suppressed as $(u/T)^{\omega\xi}$ when $\omega \xi \gg 1$.

\subsection{Long Wavelengths}
As $\omega \rightarrow 0$, the corrections to the equations from $\eta$ and $\sigma_{\textsc{q}}$ cannot be ignored.   The reason for this is simple -- as noted in Appendix \ref{app:DCanal}, at $\omega=0$ both the charge and heat currents are conserved.   Without dissipative corrections to hydrodynamics, this means that $n\tilde v = \text{constant}$ and $s\tilde v = \text{constant}$.   But $n$ and $s$ are not proportional in an inhomogeneous medium, and hence the hydrodynamic equations become ill-posed.   It is crucial to account for dissipation to resolve this discrepancy, and as noted in the main text, this leads to interesting $\omega$-dependence of localization in a disordered electron-hole plasma.

   It now becomes helpful to write the hydrodynamic equations (away from $x=x_0$) in the form \begin{equation}
\partial_x \left(\begin{array}{c} \tilde \mu \\ \tilde T \\ \tilde J \\ \tilde Q/T\end{array}\right) =  \mathsf{M}\left(\begin{array}{c} \tilde \mu \\ \tilde T \\ \tilde J \\ \tilde Q/T\end{array}\right)  + \mathbf{y} \equiv\left(\begin{array}{cc} \mathsf{0} &\ \mathsf{A} \\ \mathrm{i}\omega \mathsf{B} &\ \mathsf{0} \end{array}\right) \mathbf{\Psi}  + \mathbf{y}  \label{eq:F2eom}
\end{equation} 
where \begin{subequations}\begin{align}
\displaystyle \mathsf{A} &=  \left(\begin{array}{cc} \dfrac{\mu}{\epsilon+P} \partial_x\left(\eta\partial_x\dfrac{\mu}{\epsilon+P}\right)-\dfrac{(Ts)^2}{\sigma_{\textsc{q}}(\epsilon+P)^2}  &\ \dfrac{T^2sn}{\sigma_{\textsc{q}}(\epsilon+P)^2} + \dfrac{\mu}{\epsilon+P}\partial_x\left(\eta\partial_x\dfrac{T}{\epsilon+P}\right) \\   \dfrac{T^2sn}{\sigma_{\textsc{q}}(\epsilon+P)^2} + \dfrac{T}{\epsilon+P}\partial_x\left(\eta\partial_x\dfrac{\mu}{\epsilon+P}\right) &\ \dfrac{T}{\epsilon+P} \partial_x\left(\eta\partial_x\dfrac{T}{\epsilon+P}\right) - \dfrac{(Tn)^2}{\sigma_{\textsc{q}}(\epsilon+P)^2} \end{array}\right), \\
\mathsf{B} &= \left(\begin{array}{cc}  \dfrac{\partial n}{\partial \mu} &\  \dfrac{\partial n}{\partial T} \\  \dfrac{1}{T}\left(\dfrac{\partial \epsilon}{\partial \mu} - \mu  \dfrac{\partial n}{\partial \mu}\right) &\ \dfrac{1}{T}\left(\dfrac{\partial \epsilon}{\partial T} - \mu  \dfrac{\partial n}{\partial T}\right) \end{array}\right)=\left(\begin{array}{cc}  \dfrac{\partial n}{\partial \mu} &\  \dfrac{\partial n}{\partial T} \\ \dfrac{\partial s}{\partial \mu} &\  \dfrac{\partial s}{\partial T}  \end{array}\right),
\end{align}\end{subequations}
and
\begin{equation}
\displaystyle \mathbf{y} \equiv \left(\begin{array}{c}  \dfrac{\mu}{\epsilon+P} \partial_x\left(\eta\partial_x\dfrac{\mu \tilde J + \tilde Q}{\epsilon+P}\right) -  \tilde J \dfrac{\mu}{\epsilon+P} \partial_x\left(\eta\partial_x\dfrac{\mu}{\epsilon+P}\right)- \dfrac{\mu \tilde Q}{\epsilon+P} \partial_x\left(\eta\partial_x\dfrac{1}{\epsilon+P}\right) \\ \dfrac{T}{\epsilon+P} \partial_x\left(\eta\partial_x\dfrac{\mu \tilde J + \tilde Q}{\epsilon+P}\right) -  \tilde J \dfrac{T}{\epsilon+P} \partial_x\left(\eta\partial_x\dfrac{\mu}{\epsilon+P}\right)- \dfrac{T \tilde Q}{\epsilon+P} \partial_x\left(\eta\partial_x\dfrac{1}{\epsilon+P}\right)  \\ 0 \\ 0 \end{array}\right).
\end{equation}
Note that $\mathsf{B}$ is symmetric -- this can be shown using thermodynamic identities including (\ref{eq:Pthermo}), \begin{equation}
\epsilon + P = \mu n + Ts,
\end{equation}
and the assumption that the pressure takes the form \begin{equation}
P(\mu,T) = T^{d+1} \mathcal{F}\left(\frac{\mu}{T}\right)
\end{equation}
for some function $\mathcal{F}$, which follows from general principles for relativistic gapless fluids \cite{lucas3}.   Note that the function $\mathcal{F}$ is not completely arbitrary -- it must be chosen so that $D>0$, for example.

We have written the equations in the form (\ref{eq:F2eom}) for the following reason.   Our ultimate goal is to argue that the $\mathbf{y}$-corrections to (\ref{eq:F2eom}) are negligible as $\omega \rightarrow 0$.   Assume this to be the case.   Then the eigenvectors of $\mathsf{M}$ scale as $\sqrt{\omega}$.   The corrections $\mathbf{y}$ are all proportional to derivatives of $\tilde J$ and $\tilde Q$, which scale as $\omega$ according to (\ref{eq:F2eom}).   More precisely, accounting for $\mathbf{y}$ and using the results of Appendix \ref{app:bc}, we conclude that the local eigenvalues $\lambda$ of $\mathsf{M}$ solve the equation\footnote{Note that only the top two rows of (\ref{eq:F2eom}) are corrected by $\mathbf{y}$, and so $\mathrm{O}(\omega)$ corrections only enter here.} \begin{equation}
0 = \det \left(\lambda^2 - \mathrm{i}\omega \mathsf{AB} + \mathrm{O}(\omega,\lambda)\times \mathrm{O}(\omega)\right) \approx \det \left(\lambda^2 - \mathrm{i}\omega \mathsf{AB} \right)  \label{eq:det}
\end{equation}
and indeed at leading order all $\mathrm{O}(\omega)$ corrections to this equation are negligible.   The dominant contributions to (\ref{eq:F2eom}) come from $\mathsf{M}$ alone.    To derive (\ref{eq:det}) we have used the properties of determinants of block diagonal matrices.

We are now in a good position to approximate the solutions to (\ref{eq:F2eom}) as $\omega \rightarrow 0$.  We employ the Magnus expansion \cite{magnus}:\footnote{We neglect issues of convergence, which can be subtle.   We will see that this expansion provides the quantitatively correct result in the limit $\omega \rightarrow 0$, and will comment on its breakdown in the next subsection.}\begin{equation}
\mathbf{\Psi}(x) = \exp\left[\int\limits_0^x \mathrm{d}s \mathsf{M}(s)  + \frac{1}{2} \int\limits_0^x \mathrm{d}s\int\limits_0^s \mathrm{d}s^\prime [\mathsf{M}(s),\mathsf{M}(s^\prime)] + \cdots \right] \mathbf{\Psi}(0).  \label{eq:magnus}
\end{equation}
To capture the leading order response as $\omega \rightarrow 0$,  we need keep only the first term in (\ref{eq:magnus}).   In particular, we need to determine the eigenvalues of $\mathbb{E}[\mathsf{M}]$.   This is greatly simplified by (\ref{eq:det}), which tells us that $\lambda  = \pm \sqrt{\mathrm{i}\omega \tilde \lambda}$,  where $\tilde \lambda$ are the eigenvalues of $\mathbb{E}[\mathsf{A}]\mathbb{E}[\mathsf{B}]$.    For simplicity, let us focus on the limit where disorder is weak, and given by (\ref{eq:Fdis1}) and (\ref{eq:Fdis2}).     Since $\mathsf{AB}$ is a $2\times 2$ matrix, \begin{subequations}\label{eq:trdet}\begin{align}
\tilde \lambda _+  + \tilde\lambda_- &= \mathrm{tr}(\mathbb{E}[\mathsf{A}]\mathbb{E}[\mathsf{B}]), \\
\tilde \lambda_+ \tilde\lambda_- &= \det(\mathbb{E}[\mathsf{A}]) \det(\mathbb{E}[\mathsf{B}]).
\end{align}\end{subequations}
In the limit $u\rightarrow 0$,  it is easy to see that $\det (\mathbb{E}[\mathsf{A}]) = 0$.   Hence for small $u$, we conclude that one of the eigenvalues, which we call $\tilde \lambda_-$, is parametrically small.  From (\ref{eq:trdet}) we may approximate it by \begin{equation}
\tilde\lambda_- \approx \frac{\det(\mathbb{E}[\mathsf{A}])\det(\mathbb{E}[\mathsf{B}])}{\mathrm{tr}(\mathbb{E}[\mathsf{A}]\mathbb{E}[\mathsf{B}])}.
\end{equation}
Using thermodynamic identities for a gapless relativistic fluid \cite{lucas3}, we obtain  \begin{subequations}\label{eq:104}\begin{align}
\det(\mathsf{B}) &=  \frac{d}{T}\left(s\frac{\partial n}{\partial \mu} - n\frac{\partial s}{\partial \mu}\right), \\
\tilde\lambda_+ \approx \mathrm{tr}(\mathsf{AB}) &= \frac{1}{\sigma_{\textsc{q}}}\left[\frac{dn^2}{\epsilon+P} - \frac{\partial n}{\partial \mu} \right] = \frac{T}{\sigma_{\textsc{q}} (\epsilon+P)} \left(s\frac{\partial n}{\partial \mu} - n\frac{\partial s}{\partial \mu}\right) = \frac{1}{D} .
\end{align}\end{subequations}
We have employed (\ref{eq:Dcons}) in the last step.  Since all the disorder dependence in $\tilde\lambda_-$ comes from $\det(\mathbb{E}[\mathsf{A}])$, at leading order we may directly employ (\ref{eq:104}) when approximating $\tilde\lambda_-$.  We find after some more algebra that \begin{equation}
\tilde \lambda_- =  d\left\lbrace T^2 \left(n_0 \left(\frac{\partial s}{\partial \mu}\right)_0 - s_0 \left(\frac{\partial n}{\partial \mu}\right)_0 \right)^2 \frac{\mathbb{E}\left[(\mu-\mu_0)^2\right]}{\sigma_{\textsc{q}0}(\epsilon_0+P_0)^3} + \frac{d^2\eta_0 n_0^2}{(\epsilon_0+P_0)^3}\mathbb{E}\left[(\partial_x \mu)^2\right]\right\rbrace = \frac{d}{\tau_{\mathrm{cp}}},   \label{eq:tcp}
\end{equation}
where $\tau_{\mathrm{cp}}^{-1}$ is the momentum relaxation rate, caused by the disorder in $\mu$, in the weak disorder limit  \cite{lucas3}.\footnote{Upon employing thermodynamic identities, and using that our disorder is only in one spatial direction, our ``definition" of $\tau_{\mathrm{cp}}$ is the same as  (54) in \cite{lucas3}.}    Subscripted variables such as $\eta_0$ denote the viscosity (in this case) of the clean fluid.

Upon first glance,  it is $\tilde\lambda_+$ which dominates the response of the fluid.   However, this mode is simply the charge diffusion mode of a clean fluid (with perturbative corrections due to disorder).  This mode cannot be sourced by our injection of energy, as we emphasized in the main text.   Hence, it is the subleading $\tilde\lambda_-$ which governs the response.   We straightforwardly obtain that \begin{equation}
|\tilde J(x=0)| \sim \exp\left[- \sqrt{\frac{d\omega}{2\tau_{\mathrm{cp}}}}  x_0\right].
\end{equation}
Upon specializing to the limit where $\mu_0=0$, we obtain (\ref{eq:xiloc2}).

\subsection{Crossover Scale}
We now give a set of heuristic arguments for (\ref{omegafail}),  assuming that $\bar\mu_0=0$.     Let us begin by studying the higher order corrections in the Magnus expansion.    For simplicity, we suppose that $\xi$ is very large, so that the viscosity-dependence in $\mathsf{M}$ can be neglected.    From the schematic form of $\mathsf{M}$, the components of $\mathbb{E}[[\mathsf{M}(s),\mathsf{M}(s^\prime)]]$ and any higher order term in the Magnus expansion are at least \begin{equation*}
u^2 \left(\frac{\omega}{\sigma_{\textsc{q}}}\right)^\# \times   \text{function}(T,\xi,\ldots).
\end{equation*}
If there are an odd number of $\mathsf{M}$ in the commutator, then there is an additional factor of either $\omega$ or $\sigma_{\textsc{q}}^{-1}$.   The overall factor of $u^2$ is necessary since on the clean background, all commutators vanish and the first term in the Magnus expansion is exact.  This logic implies the perturbative Magnus expansion qualitatively fails when $\omega \sim \sigma_{\textsc{q}}$.    As (\ref{eq:rescale}) is an exact symmetry of our equations, we are forced to add the powers of $\xi$ found in (\ref{omegafail}).

We can also understand (\ref{omegafail}) by studying the breakdown of our WKB-like approximation at short wavelengths.   Keeping track of only $\sigma_{\textsc{q}}$,  which we expect is the most important dissipative correction near charge neutrality given (\ref{eq:tcp}), (\ref{eq:appc1}) should be modified to \begin{equation}
0 = -\mathrm{i}\omega \tilde n + \partial_x \left(n\tilde v - D\partial_x \tilde n + C \partial_x \tilde P\right).  \label{eq:heurg3}
\end{equation}
Recall that $D,C\sim \sigma_{\textsc{q}}$.   In fact, further modifications are necessary, since the field transformation between $\tilde \mu$ and $\tilde T$ and $\tilde n$ and $\tilde P$ leads to further complications in an inhomogeneous medium.  Still, (\ref{eq:heurg3}) suffices for our heuristic argument.   In the $\omega \rightarrow 0$ limit, we expect that $\tilde n$, $\tilde v$ and $\tilde P$ have spatial fluctuations on the order of $\xi$ -- for example, this is to enforce the (almost) conservation of charge and heat currents as $\omega \rightarrow 0$.   It is then clear that when $\omega \sim D/\xi^2 \sim  \sigma_{\textsc{q}}/\xi^2$,  the $D$ and $C$ terms in (\ref{eq:heurg3}) are no longer small,  and $\tilde n$ will not be given in terms of $\tilde v$ simply by (\ref{eq:appc1}).

\begin{figure}[t]
\centering
\includegraphics[width=5.5in]{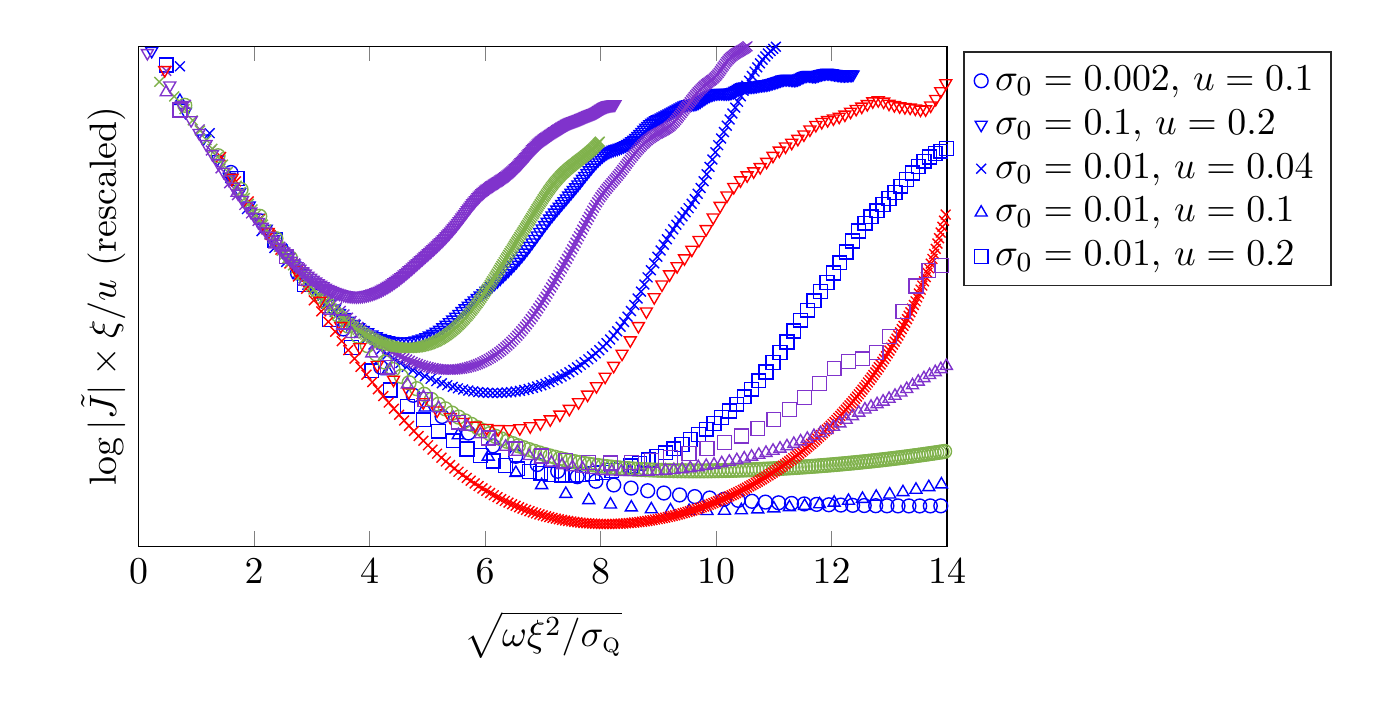}
\caption{A plot of $\log |\tilde J|$ vs. $\sqrt{\omega}$,  properly rescaled to obtain data collapse for $\omega$ smaller than (\ref{omegafail}).    Unimportant multiplicative constants have been rescaled out of $\tilde J$.    For $\omega$ larger than (\ref{omegafail}),  we see that the transition to (\ref{eq:xilocmain}) is highly sensitive to the fluid parameters.  We employed $d=2$, $\bar\mu_0=0$, $L=1000$, $\gamma=0$, $C_0=C_2=1$, and variable $\eta_0$. Different symbols represent different parameters other than $\xi$;   red represents $N=70$, blue represents $N=140$,  purple represents $N=210$, and green represents $N=280$.}
\label{fig:crossover}
\end{figure}

Figure \ref{fig:crossover} demonstrates numerically that for a broad variety of parameters,  the relaxation time scaling indeed breaks down where predicted.   Although our arguments are not rigorous, they are qualitatively justified numerically over a broad range of parameters.

  \end{appendix}
\bibliographystyle{unsrt}
\addcontentsline{toc}{section}{References}
\bibliography{soundbib}

\end{document}